\documentclass[twocolumn]{aastex631}      

\usepackage{hyperref}  
\usepackage{epsfig}
\usepackage{url}
\usepackage[maxfloats=256]{morefloats}
\def\lsi61{$\rm{LS\ I} +61^{\circ}303$}

\begin{document}    
\title{Investigation for binary characteristics of \lsi61 with optical polarization}
\author[0009-0008-3626-731X]{Jiaxin Liu}
\affiliation{Department of Astronomy, School of Physics, Huazhong University of Science and Technology, Wuhan, Hubei 430074, China}

\author{Haoyu Yuan}
\affiliation{Department of Astronomy, School of Physics, Huazhong University of Science and Technology, Wuhan, Hubei 430074, China}

\author[0009-0003-0516-5074]{Xiangli Lei}
\affiliation{Department of Astronomy, School of Physics, Huazhong University of Science and Technology, Wuhan, Hubei 430074, China}

\author{Wenlong Xu}
\affiliation{Department of Astronomy, School of Physics, Huazhong University of Science and Technology, Wuhan, Hubei 430074, China}

\author{Jumpei Takata}
\affiliation{Department of Astronomy, School of Physics, Huazhong University of Science and Technology, Wuhan, Hubei 430074, China}

\author[0000-0003-3440-1526]{Weihua Lei}
\affiliation{Department of Astronomy, School of Physics, Huazhong University of Science and Technology, Wuhan, Hubei 430074, China}

\email{jxlliu@hust.edu.cn, takata@hust.edu.cn, leiwh@hust.edu.cn}

\begin{abstract}
We investigate the optical linear polarization caused by Thomson scattering of the stellar radiation for gamma-ray binary \lsi61, which likely contains a young pulsar. Based on the pulsar binary scenario, we model the interaction between the pulsar wind and stellar wind from the massive companion star, which creates a shock. To accurately compute the resulting polarization of the stellar wind, we develop a method for the Thomson scattering that accounts for the finite size of the companion star. By fitting the optical polarization data, we constrain the system parameters, such as eccentricity, the momentum ratio of the two winds, and mass-loss rate from the companion star. We find that (i) the predicted eccentricity $e\sim 0.1$ is smaller than the values derived from the radial velocity curve and (ii) the orbital phase of the periastron is $\nu_{\rm p}=0.5-0.6$, which is consistent with the previous polarization study of Kravtsov et al. Additionally, we estimate the mass-loss rate from the companion star and the momentum ratio of two winds as $\dot{M}\sim 2\times 10^{-6}\rm M_{\odot}~{\rm year^{-1}}$ and $\eta>0.1$, respectively. Assuming that the pulsar wind carries the spin-down energy, the spin-down magnetic field of the putative pulsar inferred from these parameters is of the order of $B\sim 10^{14}\mathrm{G}$,  which may support the highly-B pulsar or magnetar scenario for the compact object of \lsi61. We also discuss the dispersion measure under the predicted orbital geometry and provide a corresponding interpretation of the pulsed radio signal detected by FAST.
\end{abstract}


\section{Introduction}
High-mass binary systems composed of the compact object (neutron star or black hole) and the massive main-sequence or evolved star have been known TeV gamma-ray sources. The TeV emitting binaries may be divided into two types; (i) so-called gamma-ray binaries \citep{2005A&A...442....1A,2013A&ARv..21...64D, 2018ApJ...867L..19A} and (ii)microquasors~\citep{2007ApJ...665L..51A,2018Natur.562...82A,2024Natur.634..557A,2024arXiv241008988L}. Gamma-ray binary systems do not exhibit the signature of the accretion process on the compact object, and its non-thermal emission extends from the radio to the TeV energy bands~\citep{2013A&ARv..21...64D}. The microqusor, on the other hand, is an accretion system, and its gamma-ray emission is probably produced by the interaction of the jet and the inter stellar matter. The population of TeV gamma-ray binaries is expected to increase with future high-energy observations.

Eight gamma-ray binary systems, PSR~B1259-63, PSR J2032+4127, LS 5039, \lsi61, 1FGL~J1018.6-5856, HESS~J0632+057, LMC P3 and 4FGL~J1405.1-6119, \citep{2013A&ARv..21...64D,2015MNRAS.451..581L,2016ApJ...829..105C,2019ApJ...884...93C} and its candidate~\citep[HESS J1832-093][]{2015MNRAS.446.1163H} have been reported in the previous observations. Two of them, PSRs B1259–63 and J2032+4127, have been firmly identified as pulsars~\citep{1992ApJ...387L..37J, 2015MNRAS.451..581L}. \lsi61 is another candidate for a pulsar system, with transient pulsed radio emissions detected by FAST observations~\citep{2022NatAs...6..698W}. Although the possibility of the black hole binary system cannot be ruled out for the gamma-ray binary systems without confirmation of the pulsed emission~\citep{2020MNRAS.498.3592M,2024A&A...683A.228J}, the pulsar binary scenario has been applied to study the non-thermal emission. In the pulsar scenario~\citep{1997ApJ...477..439T}, a relativistic pulsar wind interacts with the stellar wind/disk of the companion star, resulting in the formation of termination shocks (Figure~\ref{fig:psr}). The pulsar wind particles (electrons and positrons) are accelerated at the shock and produce the non-thermal radiation through the synchrotron radiation and inverse-Compton scattering process. The pulsar scenario has explained observed spectra and orbital modulation in the multi-wavelength bands \citep[e.g.] [and reference therein]{1999APh....10...31K, 2006A&A...456..801D, 2008A&A...488...37C, 2015MNRAS.454.1358C, 2017ApJ...836..241T, 2022A&A...658A.153C}. Although the pulsar scenario can explain the non-thermal emission of the gamma-ray binaries, further study is still necessary to identify the properties (e.g. the magnetic field and spin-down power) of putative pulsar in the gamma-ray binary systems.

The study of polarization will be another tool for investigate the gamma-ray binary system. \cite{2021ApJ...922..260X} suggest to use the X-ray polarimetry to investigate the magnetic field structure of the inter-binary shock as a result of the interaction between the pulsar wind and stellar wind. \cite{2024ApJ...974L...1K} observed the gamma-ray binary PSR B1259–63 using the Imaging X-ray Polarimetry Explorer \citep{2022JATIS...8b6002W} during an X-ray bright phase following the periastron passage in 2024 and reported detection with a polarization degree (hereafter P.D.) of 8.3\%$\pm$ 1.5\%. The observed polarization angle indicates that the magnetic field in the shock region is dominated by a component perpendicular to the shock cone axis. While X-ray polarimetry proves useful in investigating the magnetic field structure of the acceleration region, the observed X-ray flux of the known gamma-ray binaries may be insufficient to carry out further investigation, such as orbital variation, with the current detectors operated under a reasonable exposure time.  In the future, with its high sensitivity and excellent time resolution, eXTP-PFA will not only achieve higher significance but also enable us to clearly detect variations in polarization characteristics with orbital phase \citep{2025Zhou}.

In this paper, we study the linear polarization of the optical emission from gamma-ray binary systems. Polarization observations in the optical bands have been conducted for gamma-ray binaries, for example, LS~5039~\citep{2004A&A...427..959C}, HESS 0632+057~\citep{2014MNRAS.445.1761Y,2017MNRAS.464.4325Y,2021Univ....7..320M} and \lsi61~\citep{2006PASJ...58.1015N,2020A&A...643A.170K}. For LS~5039, \cite{2004A&A...427..959C} measure the linear polarization with $\sim 4.5\%$, and they do not measure significant variation of the polarization with the orbital phase. For HESS~0632+057, \cite{2017MNRAS.464.4325Y} measure the linear polarization with $\sim 1.8-2.2\%$ and find a possible orbital variation of $\sim0.7$\%~\citep{2014MNRAS.445.1761Y}. \cite{2020A&A...643A.170K} report the polarization measurement of \lsi61 and reveal the orbital variation. They find that the average P.D. of the linear polarization is $\sim 1.7-1.8$\% and its orbital profile has a double peak structure with an amplitude of $\sim 0.1$\%, suggesting the polarization source orbits with the compact object. Despite these optical observations, theoretical studies on the optical polarization of the gamma-ray binaries remain limited.

A binary system can in general produce a linear polarization of the optical emission by the Thomson scattering of free electrons in a highly ionized stellar wind and/or disk~\citep{1978A&A....68..415B, 2013ApJS..204...11H, 2022ApJ...933....5I}. \cite{2020A&A...643A.170K}  model the linear polarization of \lsi61 and constrain the orbital parameters from the observation. They find that the eccentricity ($e<0.2$) and the orbital phase of the periastron ($\nu_{\rm p}\sim 0.6$) are significantly different from $e\sim 0.5-0.7$ and $\nu_{\rm p}\sim 0.23-0.3$ determined by variations in radial velocity of the stellar atmospheric spectral lines~\citep{2005MNRAS.360.1105C,2007ApJ...656..437G,2009ApJ...698..514A}. \cite{2020A&A...643A.170K} argue that the radial velocity may be affected by the gas motion in the different parts of the Be disk. The orbital parameters proposed by \cite{2020A&A...643A.170K} are close to those proposed by modeling the orbital variation of the X-ray/TeV emission of \lsi61~\citep{2022A&A...658A.153C}.

\cite{2020A&A...643A.170K} used a simple model and assuming that a small cloud of free electrons orbits the Be star and produces the component of the linear polarization that modulates with the orbital phase. In this study, we develop a more physical model based on the pulsar binary scenario for \lsi61 and investigate the characteristics of \lsi61. We assume that \lsi61 is a colliding wind system, in which the interaction between the pulsar wind and stellar wind creates termination shocks. We take into account the linear polarization caused by the Thomson scattering of the electrons of (i) freely expanding stellar wind, (ii) shocked stellar wind and/or (iii) Be disk. In section~\ref{sec:model} and Appendix, we describe the calculation method of the optical polarization as a result of the Thomson scattering. In Section~\ref{sec:result}, we apply our model to \lsi61. In Section~\ref{sec:dis}, we discuss the implications of our results for \lsi61. Our results are summarized in Section~\ref{sec:summary}.

\section{Theoretical model of the optical polarization}
\label{sec:model}
Since \lsi61 hosts the Be star, we consider the linear polarization caused by the Thomson scattering of the free electrons in the three regions;
\begin{enumerate}
    \item freely expanding stellar wind,
    \item shocked stellar wind, and 
    \item Be disk.
\end{enumerate}

In this paper, we aim to develop a detailed model of the stellar wind component (Sections~\ref{sec:cws} and~\ref{sec:wind}). This is crucial because the observed orbital variation in optical polarization, caused by Thomson scattering of the stellar wind, can be used to diagnose the orbital parameters, as demonstrated by \cite{2020A&A...643A.170K}. Additionally, the orbital variation probably contains information about the mass-loss rate of the stellar wind and the geometry of the shock. We expect that the investigation of the Thomson scattering of the stellar wind will enhance our understanding of the nature of the gamma-ray binary systems.
For the optical polarization arose from the Thomson scattering on the Be disk, on the other hand, we refer to the previous studies done by \cite{2013ApJS..204...11H} and~\cite{2016ASPC..506..165J} (Section~\ref{sec:be}).

\subsection{Colliding wind systems}
\label{sec:cws}
\begin{figure*}
  \includegraphics[scale=0.48]{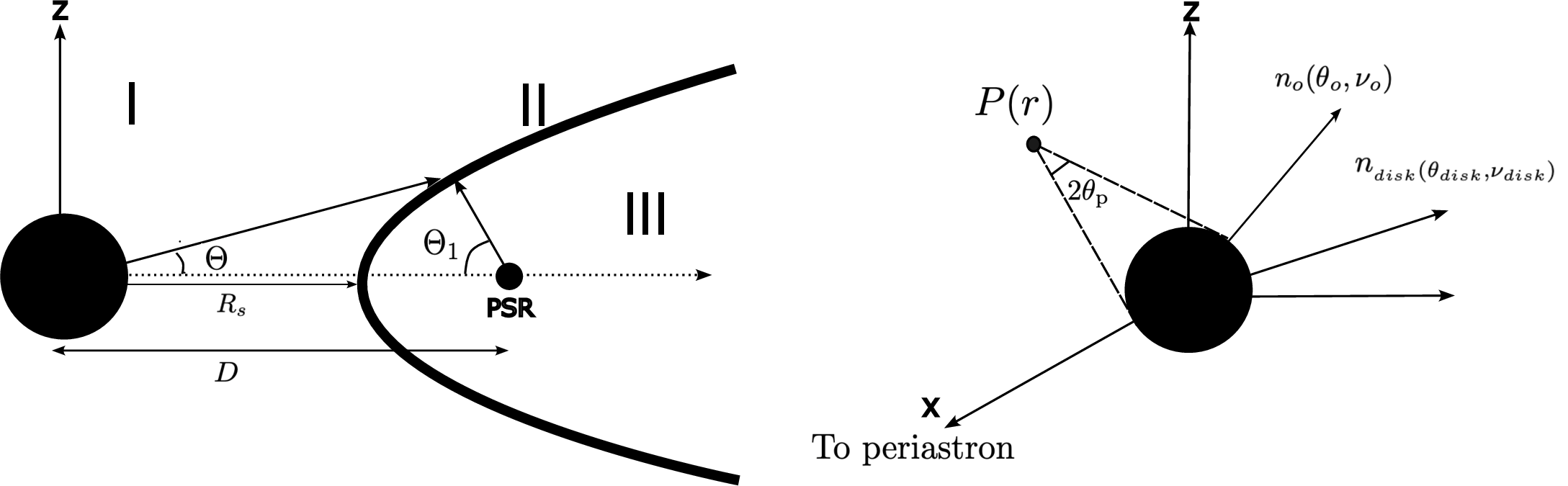}
  \caption{Left: Schematic picture of the colliding wind systems. The interaction between the pulsar wind from the pulsar (small filled circle) and outflow from the companion star (large filled circle) creates a shock (thick curved line) that wraps the pulsar. ``I'', ``II'' and ``III'' incinerates the regions of the free expanding stellar wind, the shock surface and the free expanding pulsar wind. "Z" indicates the direction of the orbital angular momentum. Right: Coordinate system applied in this study. The direction of the observer is represented by $n_{\rm o}$, where $\theta_{\rm o}$ and $\nu_{\rm o}$ are polar angle measured from Z-axis and azimuthal angle measured from X-axis, respectively. The X-axis directs toward the periastron. The $n_{disk}$ represents the axis of the Be disk that is perpendicular to the disk plane. By measuring from a scattering point, $P(r)$, the companion star convenes the sky with a solid angle of $\triangle \Omega=2\pi (1-\cos\theta_p)$, where $\theta_p$ is defined from $\cos\theta_p=\sqrt{1-R^2_*/r^2}$. }
  \label{fig:psr}
\end{figure*}

\begin{figure}
    \centering
    \includegraphics[width=1\linewidth]{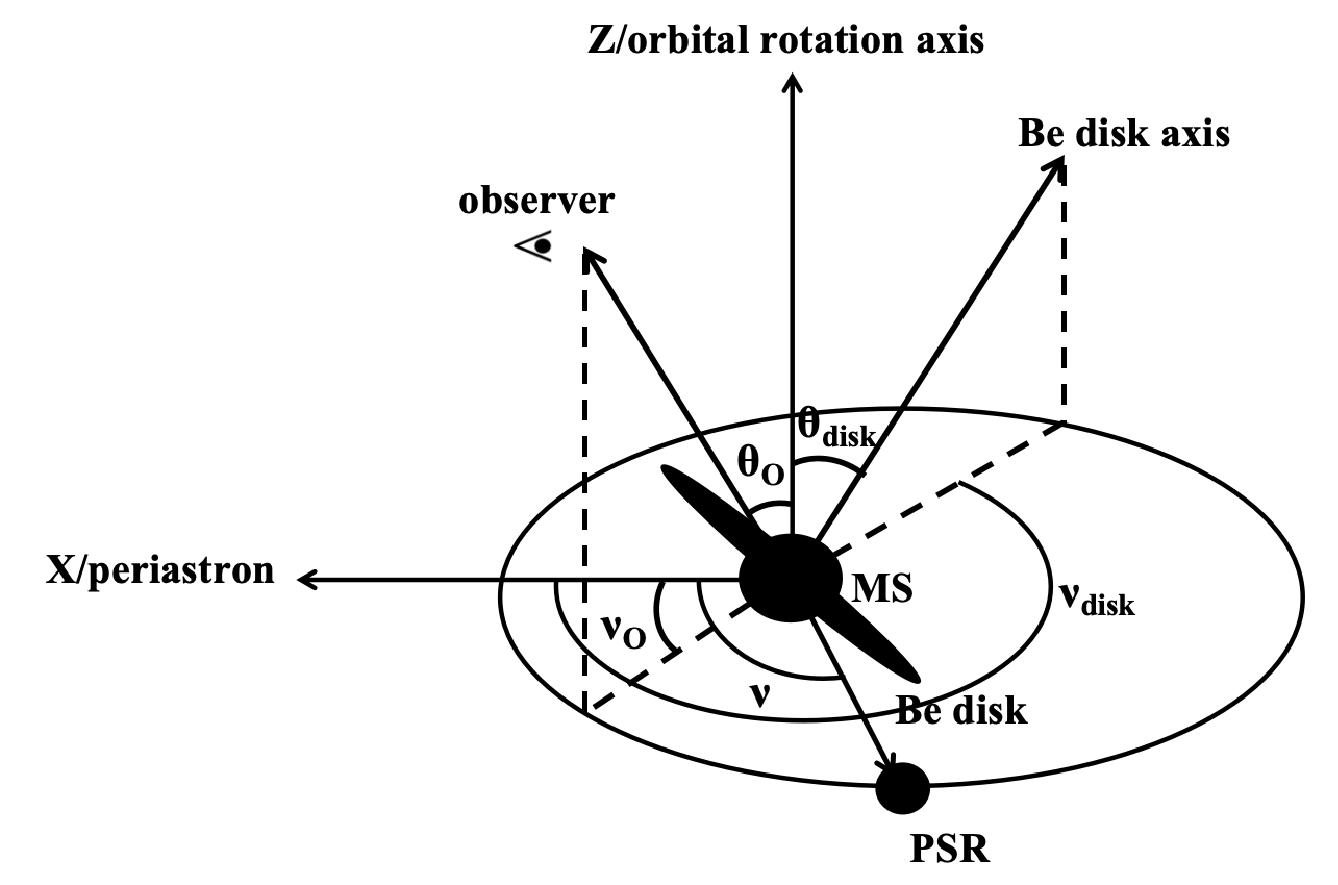}
    \caption{Schematic illustration of the geometry of the binary system. The parameter $\nu$ represents the true anomaly (angle measured from the periastron) of the pulsar.}
    \label{fig:orbital_cartoon}
\end{figure}

We assume that the interaction between the pulsar wind and the stellar outflow causes the shock cone, which produces the observed non-thermal emission in radio to TeV energy bands. We note that, except for PSRs B1259–63 and J2032+4127, the nature of the compact object and hence the emission process are still under debate. For \lsi61, for example, it is suggested that the compact object is an extremely magnetized neutron star (magnetar), and the pulsar wind is not strong enough to stop the outflow from the companion star~\citep{2012ApJ...744..106T,2012ApJ...756..188P, 2023MNRAS.525.1848S}. Consequently, a fraction of the outflow from the companion star is captured by the gravity force of the neutron star. In this magnetar scenario, the pressure of the dipole magnetic field of the neutron star prevents the matter to accrete on the neutron star surface.

In the colliding wind scenario, the location and geometry of the shock is characterized by the momentum ratio of the two winds~\citep{1996ApJ...469..729C}. For the interaction between the pulsar wind and stellar wind, the momentum ratio is given by 
\begin{equation}
 \eta\equiv \frac{L_{\rm sd}}{\dot{M}_{\rm w}v_{\rm w}c}   
 \label{eq:eta}
\end{equation}
where $L_{\rm sd}$ is the spin-down power of the pulsar, $\dot{M}_{\rm w}$ is the mass-loss rate of the stellar wind or Be disk, and $v_{\rm w}$ is the speed of the outflow. 
The spin-down powers of PSRs B1259–63 and J2032+4127 are $L_{\rm sd}\sim 8.3\times 10^{35}~{\rm erg~s^{-1}}$ and $\sim 1.5\times 10^{35}~{\rm erg~s^{-1}}$, respectively~\citep{2005AJ....129.1993M}.  For the gamma-ray binaries, the momentum ratio will be typically
\begin{eqnarray}
    \eta&=&0.025\left(\frac{L_{\rm sd}}{5\cdot 10^{36}~{\rm erg~s^{-1}}}\right) \nonumber \\
&& \times \left(\frac{\dot{M}_{\rm w}}{10^{-6}{\rm M_{\odot}}{\rm year^{-1}}}\right)^{-1}
\left(\frac{v_{\rm w}}{10^8~{\rm cm~s^{-1}}}\right)^{-1}. 
\label{eq:moment}
\end{eqnarray}
 In this study, therefore, we assume a momentum ratio of $\eta<1$, indicating the shock wraps the pulsar (Figure~\ref{fig:psr}). 

 To calculate the Thomson scattering by the free electrons in the shocked stellar wind (see section~\ref{sec:wind}), we utilize the geometry and the mass surface density of the infinitesimally thin shock as described by \cite{1996ApJ...469..729C}. Figure~\ref{fig:psr} illustrates the colliding binary system, and the distance to the apex of the shock-cone from the center of the companion star is 
\begin{equation}
    R_{\rm s}=\frac{1}{1+\eta^{1/2}}D,
\end{equation}
where $D$ is the separation between two stars. For an eccentric orbit, the distances ($D$ and $R_{\rm s}$) vary with the orbital phase. The half opening angle of the shock-cone measured from the companion star, $\Theta_{\rm s}$, is determined by
\begin{equation}
    \Theta_{\rm s}-\tan\Theta_{\rm s}=-\frac{\eta}{1-\eta}\pi.
\end{equation}
\cite{1996ApJ...469..729C} also determine by the mass-surface density, $\Sigma$, of the shocked wind as
\begin{eqnarray}
  \Sigma(\Theta,\Theta_1)&=&\frac{\dot{M}_{\rm w}}{2\pi v_{\rm w} D} \nonumber \\ 
 && \times \frac{v_{\rm w}}{v_{\rm s}(\Theta,\Theta_1)}\frac{\sin(\Theta+\Theta_1)(1-\cos\Theta)}{\sin\Theta\sin\Theta_1}
\label{eq:cdensity}
\end{eqnarray}
where $\Theta$ and $\Theta_1$ (see Figure~\ref{fig:psr} for the definition) are directions to the point on the shock surface measured from the companion star and pulsar, respectively. In addition, $v_{\rm s}$ is the speed of the shocked stellar outflow along the shock surface, and it is given by equation~(\ref{eq:vs}) in the appendix A.2. 

\subsection{Thomson scattering of stellar wind}
\label{sec:wind}
\subsubsection{Assumptions}
\label{sec:assumption}
The stellar winds from massive star is often characterized by inhomogeneous and  clumpy in the structure~\citep{2002A&A...381.1015R, 2023A&A...679A..19R}. The impact of such clumpiness on the non-thermal emission process in  the gamma-ray binary systems have been investigated in the previous studies~\citep{2013A&A...560A..32B, 2023A&A...669A..21K}. The clumpy wind will produce regions with both enhanced and reduced electron densities ($n$) compared to the isotropic case. The fraction of the volume occupation, on the other hand,   will be inversely proportional to the multiple factor representing enhancement or reduction of the electron density. 
The opposing effects of density and volume occupation fraction  tend to compensate in the optical depth.
  As a results the clumpiness may be the  minimal net impact on the overall Thomson scattering when integrated over the volume of the emission region.  Moreover, although the clumpiness may introduce a variability in the Thomson scattering with a  timescale shorter than $\sim D/v_w\sim 100$~s, the isotropic wind approximation would remain valid for interpreting the data,  which is averaged over the timescale longer than $D/v_w$. Since we will compare the calculation results with the data averaged  over several days (section~\ref{sec:lsi61}), we adopt the isotropic wind.

We anticipate that the companion star produces an isotropic and unpolarized emission.
The stellar wind from the B/O-type star will be caused by the line-driven wind and the wind reaches to the terminal speed, $v_{\rm w}\sim (2GM_{*}/R_{\rm c})^{1/2}$, at a critical distance $R_{\rm c}\sim 1.5R_{*}$ from the center of the companion star, where $M_*$ and $R_{*}$ is the stellar mass and radius, respectively~\citep{1999isw..book.....L}.
Since the critical distance is much smaller than the separation of the two stars, we assume a constant velocity of the stellar wind with the terminal speed of $v_{\rm w}\sim 6\times 10^7 (M_*/10M_{\odot})^{1/2}(R_*/10R_{\odot})^{-1/2}{\rm cm~s^{-1}}$, where $\rm M_{\odot}$ and $R_{\odot}$ are solar mass and radius, respectively. The number density of the electrons at a radial distance $r$ from the center of the star is calculated from
\begin{equation}
    n(r)=\frac{\dot{M}}{4\pi r^2 v_{\rm w}m_{\rm p}},
\label{eq:ne}    
\end{equation}
where $\dot{M}$ is the mass-loss rate of the stellar wind and $m_{\rm p}$ is the proton number density.

As described in section~\ref{sec:cws}, we consider that the interaction between the pulsar wind and stellar wind creates the shock that wraps the pulsar (Figure~\ref{fig:psr}). We assume that the shock-cone has an axisymmetric structure about the line connecting two stars. The geometry of the shock and of the down stream flow could be affected by the orbital motion. For example, the shock-cone axis may make an angle $\sim v_{\rm PSR}/v_{\rm w}$ respective to the line connecting two stars, where $v_{\rm PSR}$ is the orbital speed of the pulsar~\citep{2010A&A...516A..18D}. Since the orbital speed of the pulsar in the gamma-ray binaries will be of the order of $v_{\rm PSR}\sim  10^7~{\rm cm~s^{-1}}$ is smaller than $v_{\rm w}\sim 10^{8}~{\rm cm~s^{-1}}$ of the wind speed, the tilt of the shock-cone axis may produce a minor influence to the polarization calculation. In this study, therefore, the shock-cone axis is assumed to be the line connecting two stars.

As indicated in Figure~\ref{fig:psr}, we assume that the cavity of the stellar wind (region III) created by the shock-cone is axisymmetric along the line connecting the two stars. However, Coriolis force due to the orbital motion will induce a spiral structure in the stellar wind and its  cavity~\citep{2012A&A...544A..59B,2023A&A...677A...5K}.
For the gamma-ray binary systems, this spiral structure of the stellar wind would become significant at region beyond $(2-3)D$ from the companion star~\citep{2013A&A...551A..17Z}. Since the number density of the electrons is approximately proportional to $n_{\rm e}(r)\propto r^{-2}$ and the optical depth of the scattering is proportional to $\tau\propto r^{-1}$, we anticipate that the Thomson scattering from the spiral region will have a smaller impact on the optical polarization compared to the contribution from the inner part of the binary system. 

For the gamma-ray binary containing a Be star, the interaction between the stellar wind and pulsar wind may be obstructed at certain orbital phase by the Be disk. The scale-height of the Be disk may be described by \citep{2008ApJ...684.1374C},
\begin{equation}
H(\varpi)=H_0\left(\frac{\varpi}{R_*}\right)^\beta,
\label{eq:scale}
\end{equation}
where $H_0\sim 0.01R_*$, $\varpi$ is the axial distance from the center of the Be star, and $\beta=3/2$ for an isothermal disk. On the pulsar's orbit, the scale height is approximately 
\begin{equation}
\frac{H(D)}{D}\sim 0.05\left(\frac{D}{1{\rm AU}}\right)^{1/2}\left(\frac{R_*}{10{\rm R_{\odot}}}\right)^{-1/2}.
\end{equation}
This suggest that the pulsar interacts with the Be disk during a brief orbital phase, unless the disk plane is coplanar with the orbital plane. We therefore approximate that the stellar wind interacts with the pulsar wind throughout the entire orbit.

\subsubsection{Stokes parameters}
As indicated in Figure~\ref{fig:psr}, we divide the binary space into three region; region-I of unshocked stellar wind, region-II of the shocked stellar wind, and region-III of the cavity of the stellar wind. We introduce the Stokes parameters $(I_i, Q_i, U_i)$, where $i=\rm I$ and  $\rm II$ of the scattered radiation. Hence the {\it observed} Stoke parameters of the scattered radiation by the stellar wind becomes
\begin{equation}
I_{\rm w}=I_{\rm I}+I_{\rm II},~Q_{\rm w}=Q_{\rm I}+Q_{\rm II},~U_{\rm w}=U_{\rm I}+U_{\rm II}.
\end{equation}
It is convenient to introduce the Stokes parameters ($I_{\rm III}, Q_{\rm III}, U_{\rm III}$), which represent the contribution of the scattering if the expanding stellar wind would fill the region-III indicated in Figure~\ref{fig:psr}. We have assumed the isotropic stellar wind, suggesting net polarization will be canceled out if the stellar wind is freely expanding without the shock, namely, $Q_{\rm I}+Q_{\rm III}=0$ and $U_{\rm I}+U_{\rm III}=0$. We will calculate the observed Stokes parameters from 
\begin{equation}
I_{\rm w}=I_{\rm tot}+I_{\rm II}-I_{\rm III},~Q_{\rm w}=Q_{\rm II}-Q_{\rm III},~U_{\rm w}=U_{\rm II}-U_{\rm III},
\label{eq:stokes}
\hfuzz=1pt
\end{equation}
where $I_{\rm tot}$ is the intensity of the scattered radiation when the stellar wind is not stopped by the shock.

Calculation of the Stokes parameters are greatly simplified when the Thomson scattering is occurred in an optically thin regime. For the gamma-ray binaries, the typical mass-loss rate, wind speed and the orbital separation are $\dot{M}\sim 10^{-6}-10^{-7}{\rm M_{\odot}}~{\rm year^{-1}}$, $v_{\rm w}\sim 10^{8}~{\rm cm~s^{-1}}$ and $D\sim 1$~AU, respectively. These parameters indicate the optical depth of the order of $\tau\sim 0.0014(\dot{M}_{\rm w}/10^{-6}{\rm M_{\odot}}{\rm year^{-1}})(v_{\rm w}/10^{8}~{\rm cm~s^{-1}})(D/1{\rm AU})^{-1}$, indicating the scattering is occurred in the optically thin regime.

\subsubsection{Calculation method}
To calculate the Thomson scattering in the optically thin regime, we follow the study done by \cite{1978A&A....68..415B}. In \cite{1978A&A....68..415B}, however, the radiation source (companion star in our case) is assumed to be a point source, in which a photon from the companion star propagates in radial direction. For the gamma-ray binary systems, the size of the companion star, $R_*\sim 10{\rm R_\odot}$, can be significant fraction of the size of the system $D\sim 0.1-1$AU. In this study, therefore, we will develop the calculation method of \cite{1978A&A....68..415B} with the effect of the finite size of the companion star. Specifically, we take account the fact that by measuring from a scattering point, the companion star covers the sky with a solid angle of $\triangle \Omega=2\pi (1-\cos\theta_{\rm p})$, where $\cos\theta_{\rm p}=\sqrt{1-R^2_*/r^2}$ with $r$ being the radial distance from the companion star (right panel in Figure~\ref{fig:psr}). For each ray of the radiation coming from the stellar surface to the scattering point, the intensity is weighted by $\cos\theta_{\rm ep}$, where $\theta_{\rm ep}$ is the angle between the direction of the ray and direction perpendicular to the stellar surface. We take into account all rays illuminating the scattering point. A detailed process of the calculation is presented in Appendix~\ref{calculation}.

The intensity of Thomson scattering by the freely expanding isotropic wind, $I_{\rm tot}$ in equation~(\ref{eq:stokes}), becomes
\begin{equation}
I_{\rm tot}=I_*n_0\sigma_{\rm T}R_*\left(-\frac{\pi}{2}+2\right), 
\end{equation} 
To calculate the Thomson scattering, we introduce the direction of the observer as
\begin{equation}
\vec{n}_{\rm o} =\sin\theta_{\rm o} \cos\nu_{\rm o}\mathbf{e}_{\rm c,x}+\sin\theta_{\rm o}\sin\nu_{\rm o}\mathbf{e}_{\rm c,y}+\cos\theta_{\rm o}\mathbf{e}_{\rm c,z},
\label{eq:obs}
\end{equation}
where $\mathbf{e}_{\rm c,x}$ and $\mathbf{e}_{\rm c,z}$ represent the directions of the periastron measured from the center of mass and from the orbital axis, respectively.  
Under the assumption that the shock cone (region II) and cavity of the stellar wind (region III) are axisymmetric about the line connecting to the two stars, the Stokes parameters can be expressed as
\begin{eqnarray}
I_{i}/I_* &=& 2\tau_{{\rm c},i}+2\tau_{0,i} 
    +2\gamma_{0,i}+[\tau_{0,i}-3(\tau_0\gamma_{0})_i]\sin^2\theta_{\rm o} \nonumber\\
    && +(\tau_0\gamma_{3})_i\sin^2\theta_{\rm o}\cos2(\nu-\nu_{\rm o})  \nonumber \\
    Q_{i}/I_* &=& [\tau_{0,i}-3(\tau_0\gamma_{0})_i]\sin^2\theta_{\rm o} \nonumber \\
    && -(\tau_0\gamma_{3})_i (1+\cos^2\theta_{\rm o}) \cos2(\nu-\nu_{\rm o}) \nonumber \\
    U_{i}/I_* &=& -2(\tau_0\gamma_{3})_i \cos \theta_{\rm o}\sin2(\nu-\nu_{\rm o}),
    \label{eq:stokes1}
\end{eqnarray}
where $i={\rm II}$ or $\rm III$ and $I_*$ is the intensity of stellar radiation, and $\nu$ is the true anomaly of the pulsar (Figure~\ref{fig:orbital_cartoon}. In addition, the expressions $\tau_{\rm c}$, $\tau_0$, $\tau_0\gamma_0$, and $\tau_0\gamma_3$ that depend on the geometry of the scattering region are given by equation~(\ref{eq:taugamma}) in Appendix A.1. In equation~(\ref{eq:stokes1}), the Stokes parameters $Q_i$ and $U_i$ are measured from the direction defined by 
\begin{equation}
    \vec{y}_{\rm sky}=\vec{z}_{\rm sky}\times \vec{n}_{\rm o},
    \label{eq:vecy}
\end{equation}
where $\vec{z}_{\rm sky}$ is the orbital axis projected on the observer's sky. To compare with the model result with the observation, we introduce the parameter $\chi_0$, which is the angle between the sky North direction and $\vec{y}_{\rm sky}$, and deal it as fitting parameter (Table~1).

\subsection{Thomson scattering of Be disk}
\label{sec:be}
As reported in \cite{2020A&A...643A.170K} and~\cite{2021Univ....7..320M}, the linear polarization in the optical emission from the gamma-ray binary containing a Be star reaches $\sim 2-3\%$. We find that such a P.D. can be explained by the Thomson scattering of the stellar wind only when the mass-loss rate is unreasonably large with $\dot{M}_{\rm W}>10^{-5}{\rm M_{\odot}}{\rm year^{-1}}$. \cite{2020A&A...643A.170K}, on the other hand, suggest that the constant levels of polarization correspond to the emission from the Be disk. In this study, therefore, we also anticipate that the Thomson scattering of the Be disk contributes to the observed polarization. 

The observed emission from the disk will be a combination of the direct disk emission and the Thomson scattering of the stellar photon. It has been observed that the stellar emission ($I_*$) is one or two orders of magnitude higher than the disk emission in the optical bands, while the disk emission dominates the stellar emission in the infrared bands~\citep{1989A&A...210..295K}. Since this work studies the optical polarization, we anticipate that the observed intensity is dominated by the stellar emission. The Thomson scattering of the Be disk has been extensively investigated in previous studies~\citep[and references therein]{1996ApJ...461..828W,2013ApJS..204...11H, 2016ASPC..506..165J, 2013A&ARv..21...69R}. \cite{1996ApJ...461..828W} investigate the Thomson scattering via the Monte Carlo approaches and demonstrate that about 5-30\% of the incoming photons to the Be disk are scattered when the optical depth is $\tau\sim 1-10$. 

\cite{2013ApJS..204...11H} and~\cite{2016ASPC..506..165J}  present a detailed study of the polarization of the Thomson scattering of the Be disk. They argue that the pole on view for a perfectly axisymmetric disk is zero net polarization due to the complete cancellation of the electric vectors. As the viewing angle increases, the net polarization can be produced in the direction of the disk-axis, which is perpendicular to the disk plane, projected on the sky. They calculate the net polarization with different viewing angles, and find that predicted P.D. acquires the maximum value when the angle between the observer and the disk axis ($\theta_{\rm o,d}$) is $\theta_{\rm o,d}\sim 70^{\circ}$. The maximum P.D. when $\rho_0=10^{-10}~{\rm g~cm^{-3}}$ is about 1.5\% at the wavelength of $\sim 0.4\mu$m. We refer figure~3 in \cite{2013ApJS..204...11H} for the dependency of the P.D. on the viewing angle and base density of the Be disk.  

In our study, we do not carry out a detailed calculation for the polarization of the Be disk. Instead, we treat the intensity of the polarized emission ($I_{\rm disk}$) and the polarization angle ( $\chi_{\rm disk}$) as the fitting parameter (Table~1). We describe the polarized component of the disk emission as
\begin{equation}
    Q_{\rm disk}=I_{\rm disk}\cos2\chi_{\rm disk},~U_{\rm disk}=I_{\rm disk}\sin2\chi_{\rm disk}, 
    \label{eq:sdisk}
\end{equation}
We will diagnose the geometry of the disk from the results of the fitting~(Section~\ref{sec:hyd} and Appendix~\ref{sec:ibe}). 

The structure of the Be disk may be disturbed by its interaction with the neutron star, as discussed by \cite{2011PASJ...63..893O} and \cite{2012ApJ...750...70T} in the context of the gamma-ray binary. This disturbance has been observed as variations in the line emission profile~\citep[e.g.][]{2013A&A...559A..87Z,2015ApJ...804L..32M,2021Univ....7..320M}. For the gamma-ray binaries containing a Be star, the semi-major axis is $a\sim 0.5-10$AU, which corresponds to $D>10R_*$. On the other hand, the Thomson scattering of the Be disk is dominated by the inner region ($r<10R_*$), as illustrated in figure~6 of \cite{2013Halonen}. Consequently, the interaction may have a minor influence on the observed polarization. In our study, we assume that the polarization of the Be disk emission is time-independent.

\subsection{Fitting parameters and process}
In this study, we will compare our polarization model with B-, V- and R-band observations of \lsi61~\citep{2020A&A...643A.170K}. We calculate the Stokes parameters of the total emission as follows:
\begin{eqnarray}
    Q_{\rm tot}&=& Q_{\rm w}\cos2\chi_0-U_{\rm w}\sin2\chi_0+Q_{\rm disk}, \nonumber \\
    U_{\rm tot}&=& Q_{\rm w}\sin2\chi_0+U_{\rm w}\cos2\chi_0+U_{\rm disk},
\label{eq:stot}
\end{eqnarray}
where $Q_{\rm w}$ and $U_{\rm w}$ are calculated using equations~(\ref{eq:stokes}) and (\ref{eq:stokes1}), and $\chi_0$ is the angle of the orbital axis measured from the north direction in the sky. The degree of polarization is calculated from 
\begin{equation}
  \Pi=\frac{\sqrt{Q_{\rm tot}^2+U_{\rm tot}^2}}{I}, 
  \label{eq:pd}
\end{equation}
where the intensity is approximated as 
\[
I=I_*+I_{\rm w}+I_{\rm disk}.
\]
We note that this intensity excludes the contribution of the direct emission of and the unpolarized component of the scattered emission on the Be disk. These components are omitted because their magnitudes cannot be constrained with the optical polarization data and will be less than 10\% of the stellar emission ($I_*$). Their influence is unlikely to significantly affect the results. 

Table~\ref{tab: fitted-parameters} presents the fitting parameters of our model. Within the framework of our isotropic wind model, when the momentum ratio, $\eta$, is chosen as the fitting parameter, the mass-loss rate and the speed of the wind always appear as their ratio $\dot{M}_{\rm w}/v_{\rm w}$, which determines the number density (\ref{eq:ne}), in our calculation. Consequently, we treat $\dot{M}_{\rm w}/v_{\rm w}$ as the fitting parameter. Combined with the model, we use the Markov Chain Monte Carlo (MCMC) \footnote{\url{https://emcee.readthedocs.io/en/stable/tutorials/line/}} method to fit the data~\citep{2013PASP..125..306F}. We define the likelihood function as $\ln P(Q_m,U_m \mid \nu, \sigma, para, f) =-\frac{1}{2} \sum_{n} \left[\frac{[Q_{m}{'}(U_{m}{'}) - Q_{m}(U_{m})]^2}{s_n^2}+\ln\left(2\pi s_n^2\right) \right]$, where $\nu$ is the Orbital Phase, $para $ indicates the model parameters listed in Table~2, and $Q_{m}{'}$ (or $U_{m}{'})$ are the observed Stokes parameters and $Q_{m}$ (or $U_{m})$ are the fitted values. Additionally, $s_n^2=\sigma_n^2+f^2$ and $f$ is the model error. In our fitting process, we anticipate that $f=0$, and consequently, $s_n^2=\sigma_n^2$. 

\begin{table}
    \centering  
    \caption{Fitting parameters of polarization data}
    \begin{tabular}{c p{6cm}}
        \hline
        Parameter                     & Description and corresponding equation\\
        \hline
        $\eta$                        & Momentum ratio of two winds~(\ref{eq:eta})\\
        $\theta_{\rm o}$                 &  Inclination angle of the system~(\ref{eq:obs})\\
        $ \nu_{\rm o} $         &   True anomaly of the observer~(\ref{eq:obs})      \\
        $e$                             & Eccentricity of the system\\
        $\dot{M}/v_{\rm w}$       & Ratio of stellar mass-loss rate to the speed of the stellar wind (\ref{eq:ne})\\
        $\nu_{\rm p} $                     & Orbital phase of the periastron \\
        $\chi_0 $         & Angle between North and $\vec{y}_{\rm sky}$~(\ref{eq:vecy},\ref{eq:stot})\\
        $I_{\rm Bdisk}/I_*$           & Polarized emission from the Be disk~(\ref{eq:sdisk}) in B band\\
        $\chi_{\rm Bdisk}$  & Polarization angle of the Be disk emission~(\ref{eq:sdisk}) in B band\\
        $I_{\rm Vdisk}/I_*$           & Polarized emission from the Be disk~(\ref{eq:sdisk}) in V band\\
        $\chi_{\rm Vdisk}$  & Polarization angle of the Be disk emission~(\ref{eq:sdisk}) in V band\\
        $I_{\rm Rdisk}/I_*$           & Polarized emission from the Be disk~(\ref{eq:sdisk}) in R band\\
        $\chi_{\rm Rdisk}$  & Polarization angle of the Be disk emission~(\ref{eq:sdisk}) in R band\\
        \hline 
    \end{tabular}
    \label{tab: fitted-parameters}
\end{table}

\begin{table}
    \centering  
    \caption{Results of the fitting}
    \begin{tabular}{cc}
        \hline
        Parameter      &  Model \\
        \hline
        $\eta$         & $0.18^{+0.20}_{-0.09}$  \\ 
        $\theta_{\rm o}$     & $\pi/3$ (fixed)   \\
        $ \nu_{\rm o} $      & $1.29^{+2.76}_{-0.54}$ \\
        $e$            & $0.10^{+0.04}_{-0.04}$ \\
        $\dot{M}/v_{\rm w}$\tablenotemark{\ddag} & $1.55^{+0.30}_{-0.22}$ \\
        $\nu_{\rm p} $         &$0.54^{+0.06}_{-0.07}$ \\
        $\chi_0$         & $2.09^{+0.11}_{-0.13}$ \\
        $I_{\rm Bdisk}/I_*$  &  $1.72^{+0.01}_{-0.01}$(\%) \\
        $\chi_{\rm Bdisk}$   & $0.232^{+0.002}_{-0.002}$ \\
        $I_{\rm Vdisk}/I_*$  &  $1.78^{+0.01}_{-0.01}$(\%) \\
        $\chi_{\rm Vdisk}$   & $0.186^{+0.002}_{-0.001}$ \\
        $I_{\rm Rdisk}/I_*$  &  $1.72^{+0.01}_{-0.01}$(\%) \\
        $\chi_{\rm Rdisk}$   & $0.143^{+0.002}_{-0.002}$ \\
        \hline
    \end{tabular}
    \tablenotetext{\dag}{All angles are measured by radian.}
    \tablenotetext{\ddag}{The value is in units of $10^{12}~{\rm g~cm^{-1}}$.}
    \label{tab:fitting-i60}
\end{table}

\begin{table}
    \centering  
    \caption{Goodness of the fitting (d.o.f.=13)}
    \begin{tabular}{ccc}
        \hline
           &    Stokes Parameter Q      &  Stokes Parameter U \\
        \hline
        $\chi^2$/d.o.f. in B band  & 0.55  & 4.28   \\ 
        $\chi^2$/d.o.f.  in V band  & 0.79  & 1.05 \\
        $\chi^2$/d.o.f.  in R band  & 0.43  & 0.39 \\
        \hline
    \end{tabular}
    \label{tab:fitting indicators}
\end{table}

\section{Result}
\label{sec:result}
In this section, we present the results of our polarization study. In Section~\ref{sec:general}, we discuss the general properties of the polarization caused by the scattering of the stellar wind. In Section~\ref{sec:lsi61}, we will fit the polarization data of the LS I+61$^{\circ}$303 reported by \citep{2020A&A...643A.170K}.

\subsection{General properties of stellar wind scattering}
\label{sec:general}
\subsubsection{Influence of finite size of star} 
\label{sec:parameter} 
\begin{figure}
  \centering
  \includegraphics[width=1\linewidth]{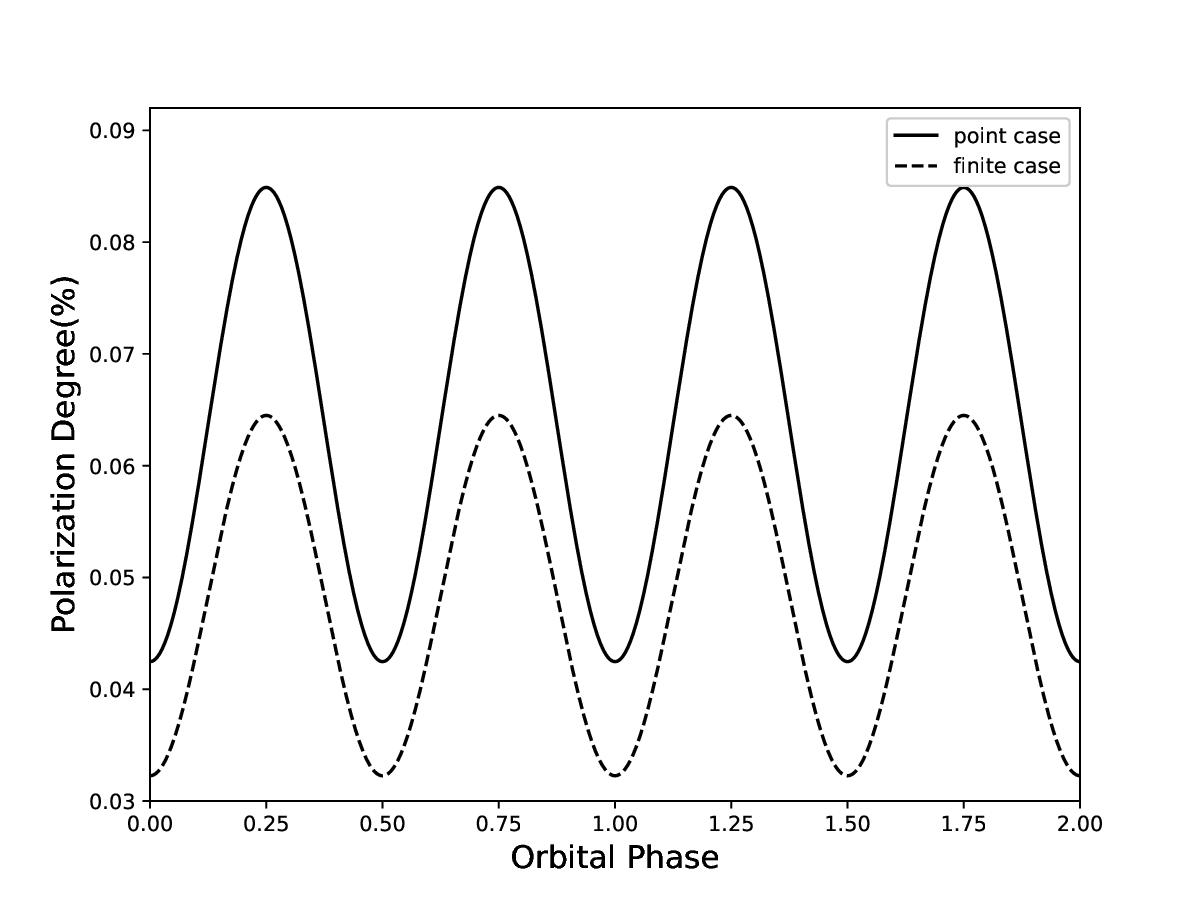}
  \caption{Comparison between the P.D.s calculated with point source approximation (dashed line) and the finite size of the companion star (solid line). The circular orbit is assumed and the phase zero correspond to the direction of the observer measured from the companion star. The other model parameters are $D=0.1$~AU, $\dot{M}/v_{\rm w}=6.7\times 10^{11}~{\rm g~cm^{-1}}$ , $\theta_{\rm o}=45^{\circ}$ and $\eta=0.25$.}
  \label{fig:point-vs-finite}
\end{figure}

Figure~\ref{fig:point-vs-finite} shows an example of the orbital modulation of the P.D. and compares between results of two calculation methods, namely the point source approximation (solid line) and the finite size (dashed line) of the star: the model parameters are the eccentricity of $e=0$, the binary separation of $D=0.1$~AU, the stellar radius of $R_*=10{\rm R_{\odot}}$ and $\dot{M}/V_{\rm W}=6.7\times 10^{11}{\rm g~cm^{-1}}(\dot{M}/10^{-6}{\rm M_{\odot}}{\rm year^{-1}})(V_{\rm W}/10^{8}{\rm cm~s^{-1}})^{-1}$.
In addition, we apply the inclination angle of $\theta_{\rm o}=45^{\circ}$ and the momentum ratio of {$\eta=0.25$}. In the figure, the zero orbital phase corresponds to {the direction of the observer}. The calculated P.D. is roughly of the order of 
\begin{eqnarray}
  \Pi&\sim &f_{\Omega}n(D)\sigma_{\rm T}D \nonumber \\
  &\sim& 0.15\% \left(\frac{f_\Omega}{0.1}\right) \left(\frac{\dot{M}/V_{\rm W}}{7\cdot 10^{11}~{\rm  g~cm^{-1}}}\right)\left(\frac{D}{0.1{\rm AU}}\right)^{-1},
\end{eqnarray}
where $n(D)=\dot{M}/4\pi m_{\rm p} D^2V_{\rm W}$ and $f_{\Omega}$ is the fraction of the stellar wind stopped by the shock.

We find in Figure~\ref{fig:point-vs-finite} that the resultant P.D. from the scattering of the stellar wind exhibits an orbital variation, displaying a double peak structure. This orbital variation arises from the change in the viewing geometry of the shock cone over the orbital cycle. The P.D. reaches the maximum value when the shock-cone axis is oriented perpendicular to the direction of the line of sight.

Figure~\ref{fig:point-vs-finite} also illustrates that the P.D. calculated considering the finite size of the star is lower compared to the point source approximation. For the finite size of the star, the rays coming from the stellar surface to a scattering point have an angular width of $\triangle \theta_{\rm p}\sim (R_*/r)$. The rays with the different incoming angles produce different P.A., leading to a depolarization effect.  

\begin{figure}
    \centering
    \includegraphics[width=1\linewidth]{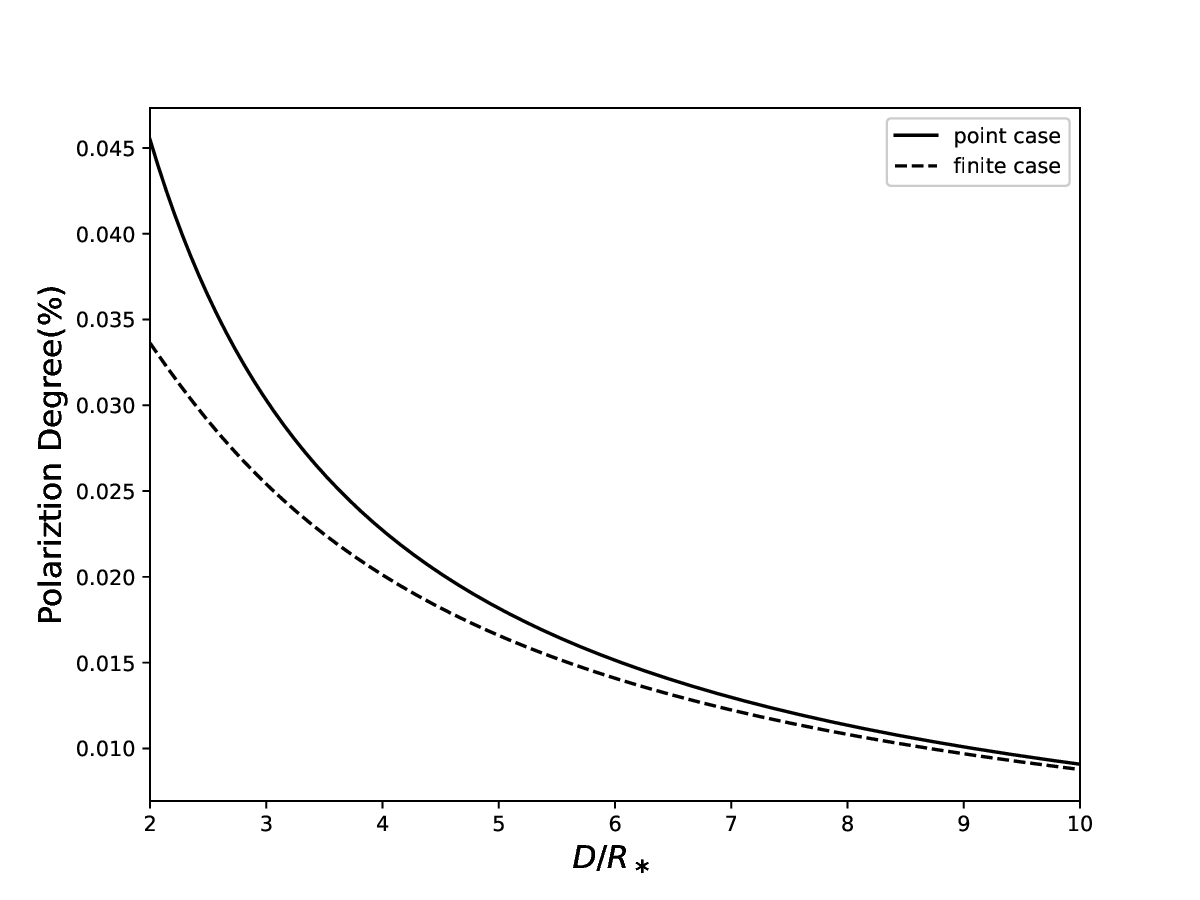}
    \caption{Dependency of the P.D.s on $D/R_*$ for point source approximation (solid line) and finite size of the radiating star (dashed line). The minimum P.D. in an orbital cycle is displayed. The line of sigh corresponds to the point case for the solid line, and the dashed line represents the finite case. When $D/R_*=2$, $PD(\%)=0.045$ for point case and $PD(\%)=0.034$ for finite case; When $D/R_*=10$, $PD(\%)=0.009$ for point case and $PD(\%)=0.0087$ for finite case}
    \label{fig:distance}
\end{figure}

Figure~\ref{fig:distance} illustrates the difference between the results of the two calculation methods as a function of the ratio of the separation ($D$) to the stellar radius ($R_*$); the figure represents the minimum P.D. in an orbital cycle. All model parameters, except for the separation of the two stars ($D$), are the same as those used in Figure~\ref{fig:point-vs-finite}. The ratio of the P.D. calculated with the point source approximation to that with the finite size of the star is {$\sim 1.3$} for $D/R_*=2$, while it is {$\sim 1$} for $D/R_*\sim 10$. We find therefore that the influence of the finite size of the star would be significant for a compact binary system, such like  LS~5039 system, which has a semi-major axis $a\sim 0.1$~AU. For large binary systems, such as LS I+61$^{\circ}$303 ($a\sim0.4$~AU) and HESS~J0632+057 ($a\sim2.3$ AU), the effect of the finite size of the star on the polarization characteristics would be negligible. Nevertheless, the following results for \lsi61 take into account the finite size of the star.

\subsubsection{Dependency on momentum ratio of two winds }
\label{sec:momentum}
\begin{figure}
    \centering
    \includegraphics[width=1\linewidth]{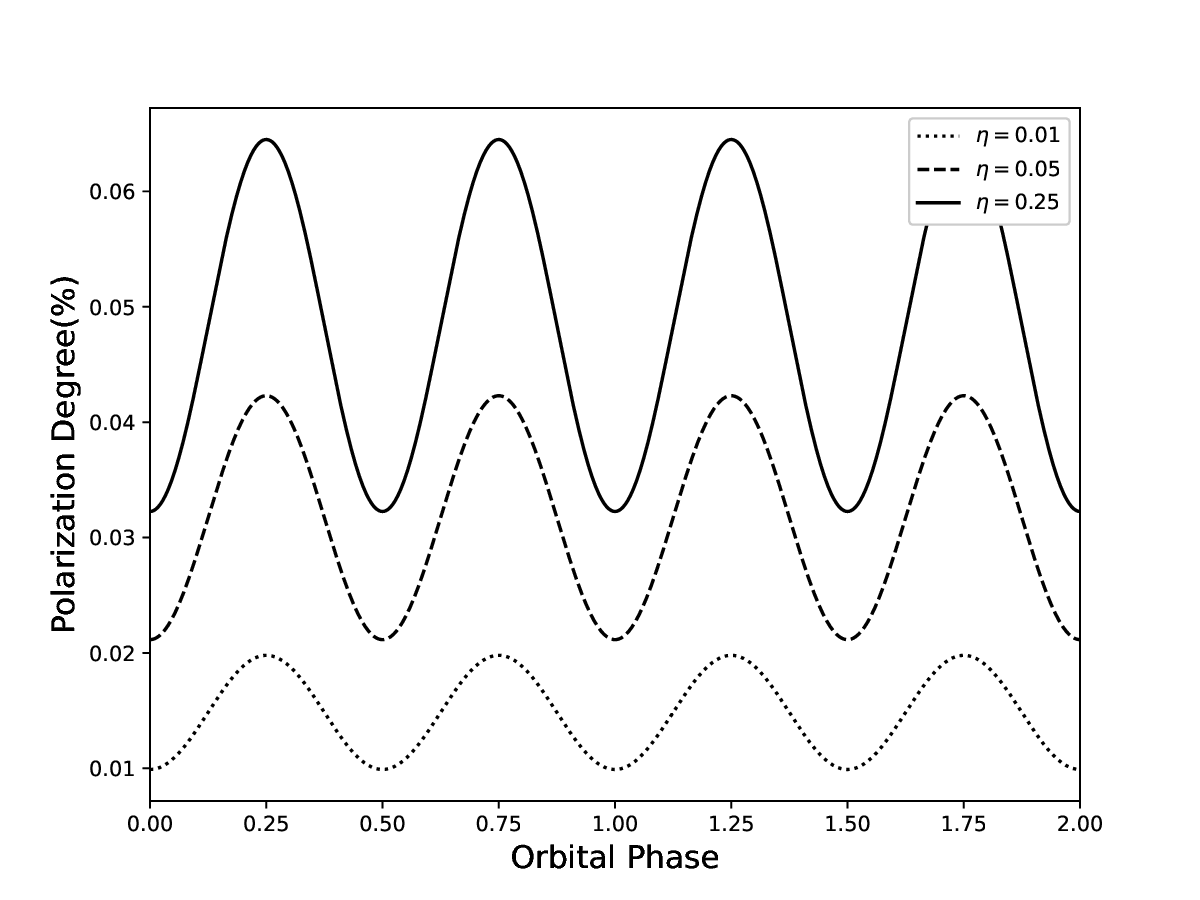}
    \caption{Orbital variation of P.D. for different momentum ratios. The dotted, solid, and dashed lines represent the cases of $\eta=0.01, 0.05$, and $0.25$, respectively. Other parameters are same as Figure~\ref{fig:point-vs-finite}.}
    \label{fig:differ-eta}
\end{figure}

\begin{figure}
    \centering
    \includegraphics[width=1\linewidth]{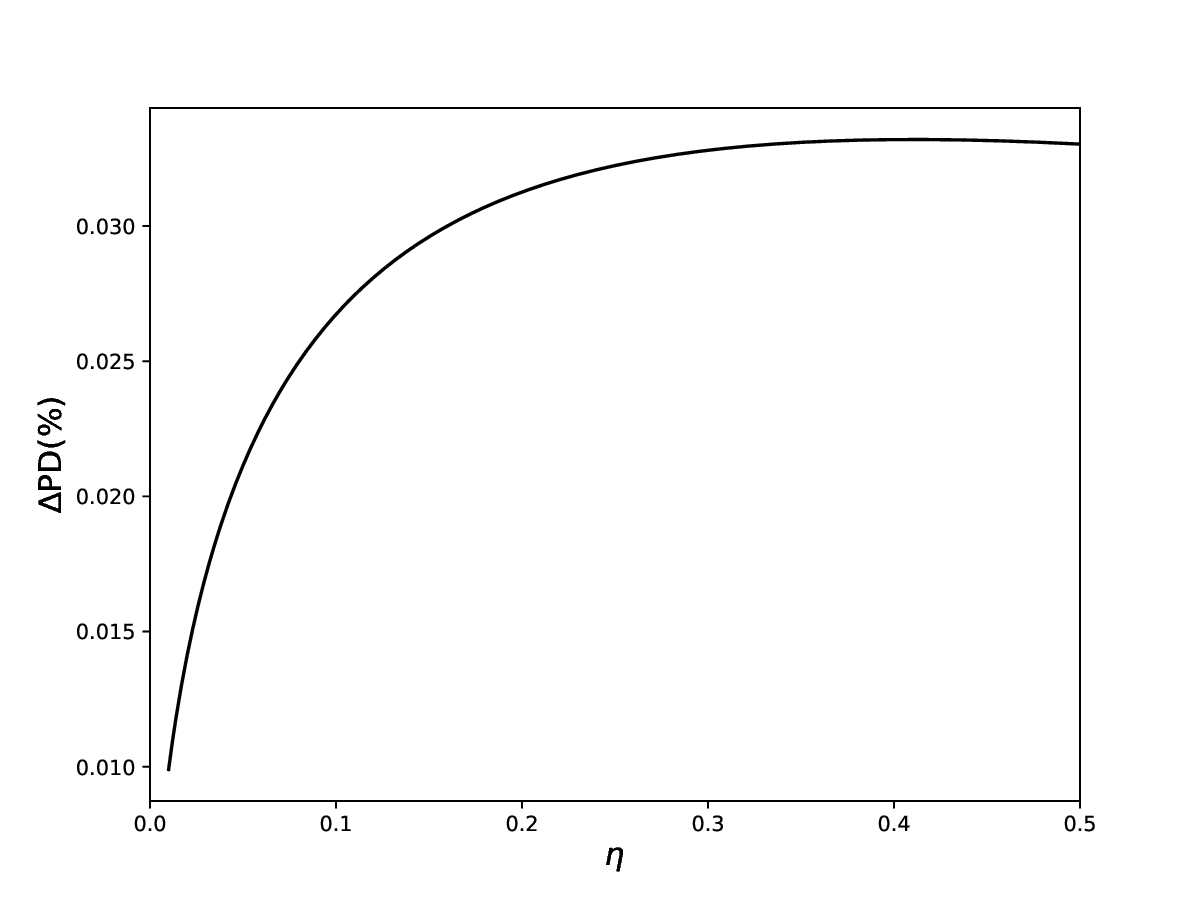}
    \caption{Amplitude of orbital profile of the calculated P.D. as a function of the momentum ratio $\eta$. Other parameters are same as Figure~\ref{fig:point-vs-finite}.}
    \label{fig:amplitude}
\end{figure}

The momentum ratio ($\eta$) of the pulsar wind to the stellar wind, as expressed by equation~(\ref{eq:moment}), is one of the important parameters, in the context of the gamma-ray binary, since it determines the size of the shock region and the characteristics of the emission from the shock region. Figure~\ref{fig:differ-eta} illustrates the dependency of the P.D. on the Thomson scattering of the stellar wind: the other model parameters are the same as those used in Figure~\ref{fig:point-vs-finite}. Figure~\ref{fig:amplitude} shows the amplitude of the orbital variation as a function of the momentum ratio.

We can find in Figure~\ref{fig:differ-eta} that as the momentum ratio decreases (i.e. as the pulsar's spin down power relative to the stellar wind weakens), the predicted P.D. decreases. This occurs because a lower value of the momentum ratio results in a narrower opening angle of the shock cone wrapping the pulsar and reducing the amount of the stellar wind stopped by the shock. As a result, the Thomson scattering of the stellar wind produces a lower P.D.. Additionally, as shown in Figure~\ref{fig:amplitude}, the orbital amplitude of the P.D. also decreases with decreasing momentum ratio; in the figure, $\triangle$P.D.$\sim 0.03$ for $\eta=0.25$ and $\triangle$P.D.$\sim 0.01$ for $\eta=0.01$. The figure also show that the amplitude is less dependent the momentum ratio when $\eta>0.1$. These dependencies on the momentum ratio may be utilized to diagnose the magnitude of the $\eta$ for a specific binary system by the fitting the observed orbital modulation of the P.D (section~\ref{sec:lsi61}). 

\subsubsection{Dependency on orbital eccentricity}
\begin{figure}
  \centering
  \includegraphics[width=1\linewidth]{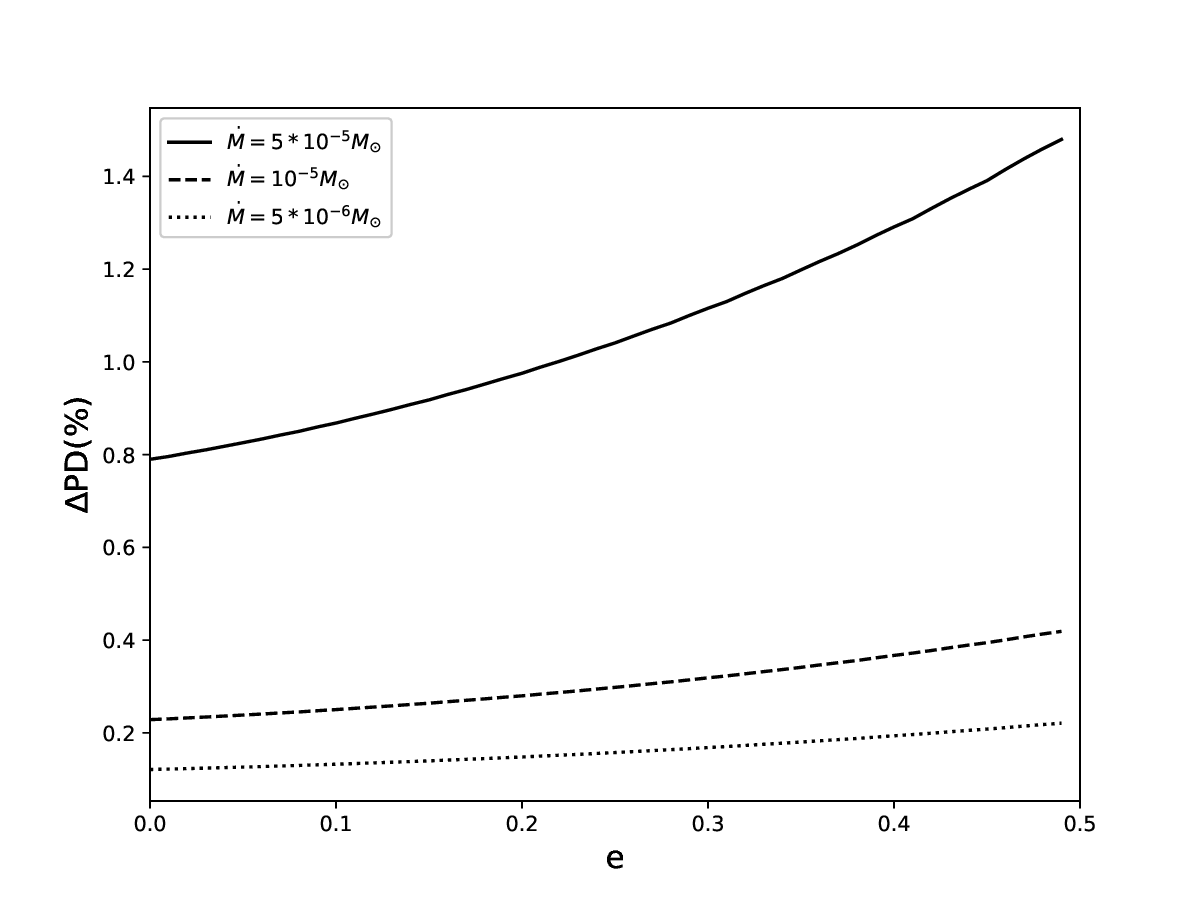}
  \caption{Amplitude of the orbital variation of the P.D. as a function of the eccentricity $e$. The dashed line, solid line and dotted line are results for $\dot{M}=5 \times 10^{-5}{\rm M_{\odot}}{\rm year^{-1}}$, $10^{-5}{\rm M_{\odot}}$ and $5\times 10^{-6}{\rm M_{\odot}}$, respectively. The parameters are $a=0.14$AU, $v_{\rm w}=10^8~{\rm cm~s^{-1}}$ and $\eta=0.25$. The discretion of the observer is assumed to be $\theta_{\rm o}=45^{\circ}$ and $\phi_{\rm o}=0$. }
  \label{fig:ePD}
\end{figure}

As one of the orbital parameters, eccentricity has a significant impact on the orbital modulation of polarization. Figure~\ref{fig:ePD} illustrates the dependency of the orbital amplitude of the P.D. on the eccentricity:semi-major axis is fixed to $a = 0.14~$AU and mass-loss rate is assumed to be $\dot{M}=5\times 10^{-5}{\rm M_{\odot}}{\rm year^{-1}}$ (solid line),  $10^{-5}{\rm M_{\odot}}{\rm year^{-1}}$ (dashed line) and $5 \times 10^{-6}{\rm M_{\odot}}{\rm year^{-1}}$ (dotted line). The figure shows that as the eccentricity increases, the amplitude also increases. Additionally, we note that the orbital elongation associated with eccentricity alters the overall morphology of polarization variation. Specifically, the symmetric double-peaked structure seen in Figure~\ref{fig:differ-eta} becomes less distinct with increasing eccentricity. In our model, therefore, the observed shape of the polarization modulation is influenced by the degree of orbital eccentricity.

Figure~\ref{fig:ePD} demonstrates that the amplitude of the P.D. increases with ratio of the mass-loss rate and speed of stellar wind, $\dot{M}/v_{\rm w}$, if the momentum ratio ($\eta$) is fixed. This trend arises because a higher   $\dot{M}/v_{\rm w}$ leads to a higher electron number density of the stellar wind, resulting in a greater P.D. averaged over the one orbit. Consequently, the information of the observed orbital variation of the P.D. provides the formation of $\dot{M}/v_{\rm w}$ of the stellar wind.

\begin{figure}
  \centering
  \includegraphics[width=1\linewidth]{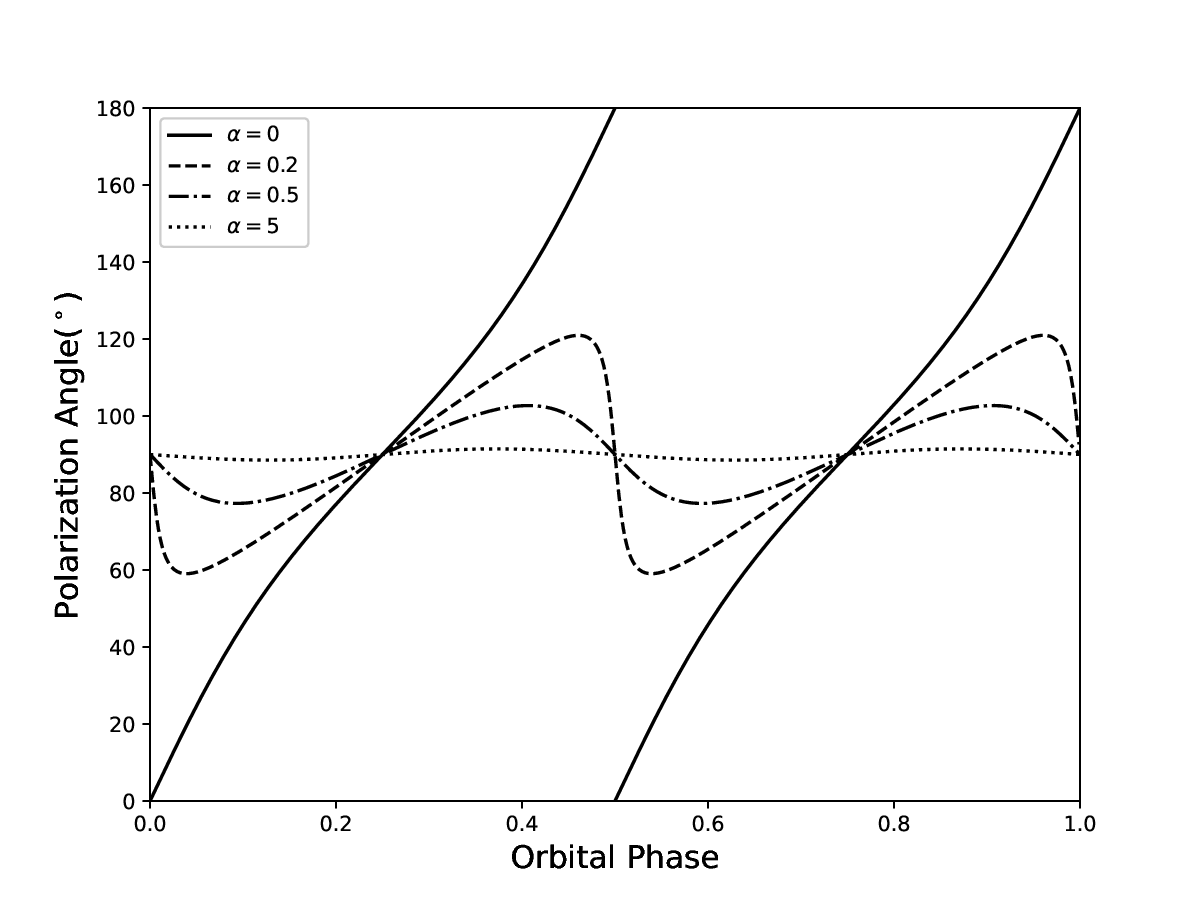}
  \caption{Orbital variation of the P.A. The parameter $\alpha$ represents the ratio of the polarized emission from the Be disk to the shocked stellar wind. For $\alpha=0$, the emission arises solely from the shocked stellar wind. The P.A. of the Be disk component is assumed to be $90^\circ$. The eccentricity is $e=0$  and  the phase zero is defined as the direction pointing toward the observer as measured from the companion star.}
  \label{fig:phi_chi}
\end{figure}

\subsubsection{Orbital variation of P.A.}
As outlined in Section~\ref{sec:wind}, we assume an axisymmetric structure of the shocked stellar wind about the line connecting between two stars. The emission resulting from the Thomson scattering integrated over the shock-cone surface is polarized in the direction of the shock-cone axis projected on the sky. Consequently, when the observed emission arises solely from the shocked stellar wind, the observed P.A. will rotate by $360^{\circ}$ during one orbital period, as illustrated by the solid line in Figure~\ref{fig:phi_chi}.

In the case of \lsi61, the observed emission arises from the combination of the two components; (i) the stellar wind and (ii) the Be disk. Figure~\ref{fig:phi_chi} illustrates how the orbital variation of the P.A. depends on the ratio ($\alpha$) between the intensity of the polarized component of the Be disk and that of the stellar wind; the results assume a circular orbit, with the phase zero defined as the direction pointing toward the observer as measured from the companion star. In addition, the P.A. of the Be disk component is steady at the magnitude of $90^\circ$. As shown in Figure~\ref{fig:phi_chi}, increasing the relative contribution of the Be disk emission leads to a reduction in the amplitude of the orbital variation of P.A.

\subsection{Application to \lsi61}
\label{sec:lsi61}
\begin{figure*}
    \centering
    \includegraphics[width=1\linewidth]{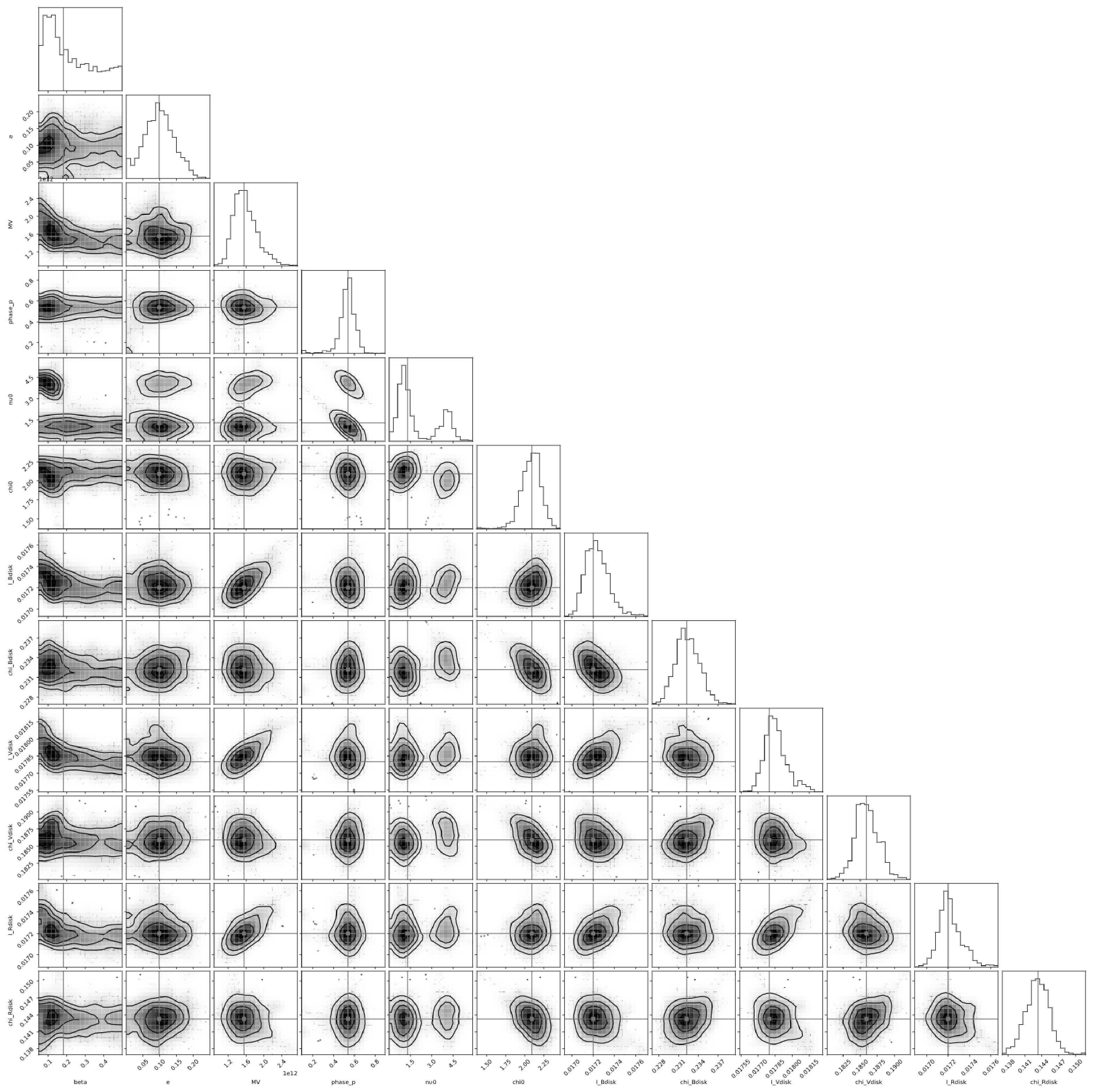}
    \caption{Posterior distributions for parameters of model described in Sect.2. Diagonal panels: distributions of model parameters from Table~\ref{tab:fitting-i60}. Lower-triangle panels: joint posterior distributions of two parameters.}
    \label{fig:corner_i60}
\end{figure*}

\begin{figure}
    \centering
    \includegraphics[width=1\linewidth]{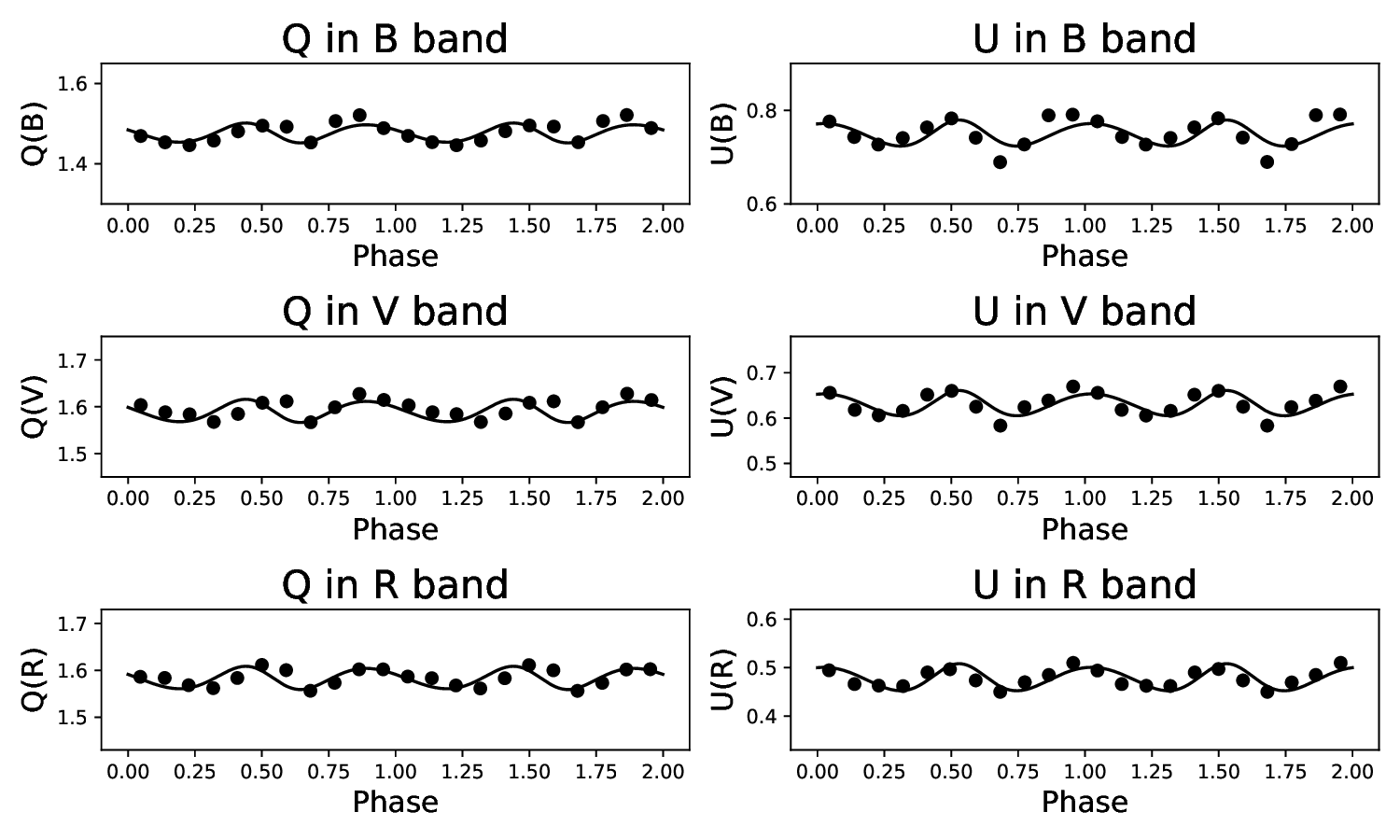}
    \caption{The best-fit curves corresponding to Table~\ref{tab:fitting-i60}. The panels correspond to the Stokes parameters $q$ and $u$ (from left to right), with the orbital phase shown along the horizontal axis, and the vertical axes (from top to bottom) correspond to the BVR band. The circle points are the observational data from Kravtsov et al. (2020), and the curves represent the fitted results.}
    \label{fig:fitting-i60}
\end{figure}

\begin{figure}
    \centering
    \includegraphics[width=1\linewidth]{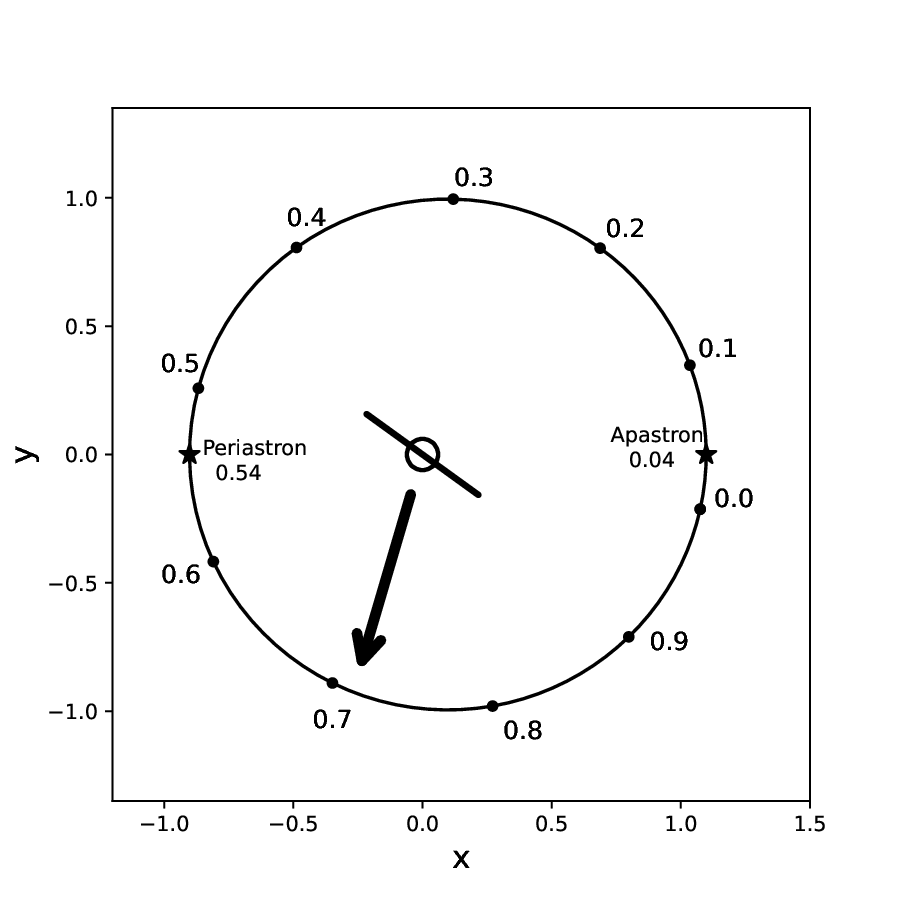}
    \caption{Orbital geometry of \lsi61 predicted by the optical polarization (Table~\ref{tab:fitting-i60}). The stars represent the periastron (0.54)
      and apastron (0.04). the black dots on the ellipse are spaced by $\triangle \phi$ = 0.1. The black arrow indicates the direction toward the observer. The black solid line represents the line of nodes between the orbital plane and the Be disk (calculation details are given in Appendix B)}
    \label{fig:fitting-i60-orbit}
\end{figure}

We apply our model to fit the observation of the B-, V-, and R-bands of \lsi61. Tables~1 and~2 summarize the fitting parameters and the corresponding results, respectively, while Figures~\ref{fig:corner_i60} and \ref{fig:fitting-i60} present the corner plot and the best fit curve, respectively. Additionally, Figure~\ref{fig:fitting-i60-orbit} illustrates the orbital geometry predicted by the fitting result.  

For the result presented in Table~2, we fix the system inclination angle $\theta_{\rm o}$. If we fit the inclination angle, the fitting procedure suggests a preferred angle of $\theta_{\rm o}\sim 90^\circ$ (i.e. edge-on view), which is consistent with the result of \cite{2020A&A...643A.170K}. Meanwhile, the X-ray and gamma-ray emissions, which are likely originated from the inter-binary shock, do not exhibit an eclipse feature, which might be expected under the edge-on geometry. Previous studies have not generally favor the edge-on view, although the angles are poorly constrained. \cite{2006PASJ...58.1015N} suggested that the inclination angle $i<30^{\circ}$. Under the constraint that the mass of the optical component of the system is in the range of $10-15{\rm M_\odot}$, \cite{2005casares} suggested that the compact object would be a neutron star for inclinations $25^\circ<i< 60^\circ$, and based on radial velocity measurements, they pointed to a rough upper limit of $i<60^\circ$. In this study, therefore, we adopt a fixed inclination angle of $\theta_0=\pi/3$. We note that the fitting results are not sensitive to the exact choice of this angle. Table~\ref{tab:fitting indicators} provides the reduced chi square, $\chi^2$/d.o.f., of the fitting for each Stokes parameter and each bands;  the degree of freedom (d.o.f.=13) is calculated from ``number of the data point - number of the parameter''.

The first column of the corner plot (Figure~\ref{fig:corner_i60}) indicates that the momentum ratio ($\eta$) is only loosely constrained by $\eta>0.1$.  This trend arises because the momentum ratio mainly influences the orbital amplitude of the P.D., as discussed in section~\ref{sec:momentum}, and for $\eta>0.1$, the amplitude becomes less sensitive to the momentum ratio, as demonstrated in Figure~\ref{fig:amplitude}. We find, on the other hand, that the fitting disfavors a momentum ratio of $\eta\ll 0.1$, which predicts an orbital amplitude of the P.D. significantly smaller than the observed one.  

The fourth column of the corner plot (Figure~\ref{fig:corner_i60}) indicates two potential solutions for the true anomaly of the observer ($\nu_{\rm o}$, see Figure~\ref{fig:orbital_cartoon} for the definition); either $\theta_{\rm o}\sim 1.3$~rad or 4.3~rad. These two solutions differ by approximately $\pi$~rad and reflect the degeneracy in the current model of the Stokes parameters $Q_i$ and $U_i$ introduced by the equation~(\ref{eq:qui}). We adopt the solution of $\nu_{\rm o}<\pi/2$ in Table~2 and Figure~\ref{fig:fitting-i60-orbit}, for consistently with geometry proposed by the previous studies~\citep[e.g.][]{2005MNRAS.360.1105C,2007ApJ...656..437G}. 

As shown in the Table~\ref{tab:fitting indicators}, the reduced chi-square values, $\chi^2$/d.o.f.,  presented in the table indicate that the fittings in the V and R bands are generally accepted. In contrary, the $\chi^2$d.o.f. for the Stokes parameter U in the B band is significantly higher (4.28), which implies a notable deviation between the model and the data. This discrepancy  would indicate the limitation of the current model.  For instance, we assume a steady Be disk structure unaffected by the interaction with the neutron star. However, periodic disturbances caused by the neutron star could influence the orbital modulation of polarization, particularly since Thomson scattering in the Be disk dominates the polarized emission. A more sophisticated model that accounts for disk variability and interaction effects may be required to fully interpret the observations. 

Our results generally consistent with the original study done by \cite{2020A&A...643A.170K}; (i) the predicted eccentricity $e\sim 0.1$ is smaller than values derived by \cite{2005MNRAS.360.1105C} $\sim 0.72$ and \cite{2007ApJ...656..437G} $\sim 0.55$ and (ii) orbital phase of the periastron is $\nu_{\rm p}=0.5-0.6$~ \citep[0.62 in][]{2020A&A...643A.170K}. We note that the FAST observation confirm the transient pulsed radio emission at the phase of $\sim 0.5-0.6$~\citep{2022NatAs...6..698W}, which corresponds to a phase close to the periastron for our orbital solution. 

Finally, the result of the fitting suggests that the contribution of the linearly polarized emission from the Be disk is of the order of $I_{\rm disk}/I_*\sim 1.7$~\%, as presented in Table~2. Previous studies suggest that such a high P.D. can be produced by the Be disk with base mass density of $\rho_0\sim 10^{-10}-10^{-9}~{\rm g~cm^{-3}}$ and the inclination angle of $\theta_{\rm o,d}\sim 70^{\circ}$ from the direction of the observer \citep[see figure~3 of][]{2013ApJS..204...11H}. By taking $\theta_{\rm o,d}=70^{\circ}$ and by assuming that the direction of the polarization repents the direction of the disk axis projected on the sky, we obtain $\theta_{\rm disk} \sim 127.5^{\circ}$ and $\phi_{\rm node} \sim 143.97^{\circ}$, the solid line in Figure~\ref{fig:fitting-i60-orbit} represents the node line between the orbital phase and the Be disk (see Appendix~\ref{sec:ibe} for detailed calculation). It has been observed an enhancement of the X-ray and TeV emission during the orbital phase of $\sim 0.4-0.8$~\citep{2009ApJ...706L..27A,2017MNRAS.470.1718C}. With the proposed orbital geometry, the enhancement happens when the first Be disk/PSR interaction and the periastron passage, as indicated by Figure~\ref{fig:fitting-i60-orbit}.

\section{Discussion}
\label{sec:dis}
\subsection{Compact object in \lsi61}
The nature of the compact object of \lsi61 is a long-standing issue since its identification of a possible gamma-ray emission~\citep{1978Natur.272..704G}. Due to the uncertainties of the compact object, its emission process has been discussed based on the microquasar scenario \citep[e.g.][]{2004A&A...425.1069B,2017MNRAS.468.3689M,2024A&A...683A.228J} or the pulsar binary scenario~\citep[e.g.][]{1981MNRAS.194P...1M, 2006A&A...456..801D, 2022A&A...658A.153C}. \cite{2022NatAs...6..698W} report a transitional pulsed radio signal from \lsi61 using the FAST observation carried out in January 2020 and its period of $P\sim 0.26$~s, suggesting the compact object is a young pulsar. Within the framework of the pulsar binary scenario, the emission process of \lsi61 has been discussed by two scenarios, namely the binary system in propeller phase~\citep{2012ApJ...756..188P,2022ApJ...940..128S} or the colliding-wind system~\citep{1981MNRAS.194P...1M, 2022A&A...658A.153C} as described in the current study. \cite{2012ApJ...744..106T} report on the \verb|Swift| detection of the magnetar-like X-ray flare from the direction of \lsi61, suggesting that the compact object is a highly magnetized neutron star. \cite{2012ApJ...756..188P} develop the magnetar scenario of \lsi61 and suggest that the neutron star undergoes a transition between a propeller phase around the periastron and the ejector phase around the apastron. \cite{2022ApJ...940..128S} carry out a further study of the magnetar scenario of \lsi61 and conclude that a propeller torque is required to explain the possible spin period observed by FAST.

As described in section~\ref{sec:lsi61} and Table~\ref{tab:fitting-i60}, our results predict that the ratio of the mass-loss rate to the wind speed is of the order of $\dot{M}_{\rm w}/v_{\rm w}\sim (1.55)\times 10^{12}~{\rm g~cm^{-1}}$. With a typical wind speed of $v_{\rm w}=10^8~{\rm cm~s^{-1}}$ from the companion star, the mass-loss rate is estimated to be $\dot{M}_{\rm w}\sim 2\times 10^{-6}{\rm M_{\odot}}~{\rm year^{-1}}$. Our model also favors $\eta\ge 0.1$ for the momentum ratio of the two winds. If the pulsar wind carries a substantial fraction of the spin-down power given by  
\begin{equation}
    L_{\rm sd}=\frac{(2\pi)^4B_{\rm NS}^2 R_{\rm NS}^6}{6c^3P_{\rm NS}^4},
\end{equation}
where $B_{\rm NS}$ and $R_{\rm NS}=10^6$~cm are the magnetic field and the radius of the neutron star, respectively. The magnetic field is estimated to be \begin{eqnarray}
    B_{\rm NS}&\sim& 1.5\times 10^{14}{\rm G}\left(\frac{\eta}{0.1}\right)^{1/2}
    \left(\frac{P_{\rm NS}}{0.26~{\rm s}}\right)^{2} \nonumber \\
    &&\left(\frac{\dot{M}/v_{\rm w}}{1.55\cdot10^{12}~{g~\rm cm^{-1}}}\right)^{1/2}\left(\frac{v_{\rm w}}{10^{8}~\rm{cm~s^{-1}}}\right).
 \label{bfield}
\end{eqnarray}
The  MCMC fitting yields a momentum ratio of $\eta=0.18^{+0.20}_{-0.09}$. The uncertainty of the momentum ratio  introduces a factor of $\sim 2$ variation in the magnetic field estimation. In contrast, the wind\
  parameter $\dot{M}/v_w=1.55^{+0.30}_{-0.22}\times 10^{12}~{\rm g~cm^{-1}}$ is well constrained by the fit, and its relatively small  uncertainty has a minor impact on the magnetic field estimation.
  In this  study,  the dominant source of uncertainty in the magnetic field estimation arises from the stellar wind speed.   Previous studies  measure or estimate  wind velocity  in  a range of  $v_w\sim 0.1-2\times 10^{8}~{\rm cm}$~\citep{1988Waters,1981Snow, 1989Prinja,1990ApJ...361..607P}. For instance  \cite{1989Prinja} analyze  the P-Cygni profile of $40$ Be stars and derive the  wind speeds. As a result, the magnetic field strength derived from  equation~(\ref{bfield}) carries an uncertainty  up to a factor of ten. Despite this, our calculation suggests that the magnetic field likely exceeds  $B_{NS}>10^{13}$~G. Therefore, our polarization study based on the colliding wind model may support the presence of either  a high-B pulsar, which is an  rotation powered pulsars exhibiting  magnetar-like outburst~\citep{2016ApJ...829L..21A,2024ApJ...976...56S}, or a magnetar scenario for
  \lsi61~\citep{2012ApJ...744..106T}.

\begin{figure}
\includegraphics[scale=0.5]{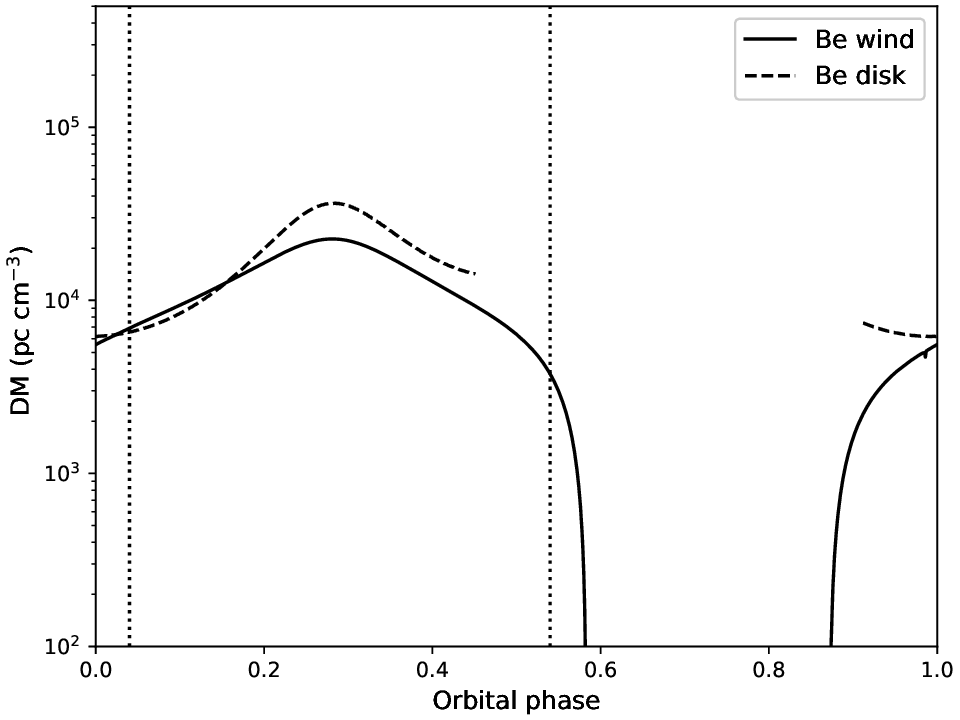}
  \caption{Predicted DM  for the stellar wind (solid line) and of Be disk (dashed line) of \lsi61 along the line-of-sight measured from the position of the pulsar. The vertical dotted lines around the orbital phases $\sim 0.55$ and $\sim 0.05$ are the positions of the periastron and apastron, respectively.}
  \label{fig:nh}
\end{figure}

\subsection{Electron column density}
\label{sec:hyd}
\cite{2022NatAs...6..698W} report a pulsed radio signal from \lsi61 using the FAST observation conducted in January 2020, which corresponds to the orbital phase of $\sim 0.6$. However, subsequent observations have not measured the pulsed radio emission, suggesting the transient nature of the radio emission. These FAST observation rises unresolved questions; (i) origin of the transient behavior and (ii) the reason for its occurrence specifically at orbital phase of $\sim 0.6$. \cite{2021A&A...652A..39C} studies the electron column density for the gamma-ray binary PSR~B1259-63 and conclude that the pulsed radio emission from the pulsar at the periastron passage is obscured by the Be disk due to free-free absorption. This study would suggest that the influence of the stellar wind or Be disk is a more probable reason for the transient nature of the radio emission of \lsi61. To investigate the possibility of observation of the pulsed radio emission from the putative pulsar, therefore,  it may be worthwhile to study the dispersion measure (hereafter DM) predicted by the proposed orbital geometry (Table~2 and Figure~\ref{fig:fitting-i60-orbit}).

We calculate the electron number density of the unshocked wind and the column density of the shocked wind from equations~(\ref{eq:ne}) and (\ref{eq:cdensity}), respectively. For the Be disk, we may describe the mass density distribution as~\citep{2008ApJ...684.1374C}
\begin{equation}
  \rho_{\rm d}(\varpi,z)=\rho_0\left(\frac{R_*}{\varpi}\right)^{n}{\rm exp}\left(-\frac{z^2}{2H^2}\right),
\label{eq:density}
\end{equation}
where $\rho_0$ is the mass density at the mid-plane of the disk at the stellar surface and $n=3.5$ is for the isothermal disk. We calculate the free-electron density from $\rho_{\rm d}/m_{\rm p}$. For the base density, we apply $\rho_0=10^{-10}~{\rm g~cm^{-3}}$ as estimated in section~\ref{sec:lsi61}.
 We trace the trajectory from the pulsar to the observer and we anticipate that the contribution of the pulsar wind region to the DM is negligible.

 Figure~\ref{fig:nh} presents the calculated DM, where the solid and dashed lines indicate contributions from the stellar wind and the Be disk, respectively, along the line of sight. We find in the figure that the calculated DM is of the order of DM$\sim 10^4~{\rm pc~cm^{-3}}$ for most of the orbital phase. This DM is substantially larger than $240.1~{\rm pc~cm^{-3}}$ measured during the 2020 FAST observation~\citep{2022NatAs...6..698W}. Moreover, the figure also suggests a potential window of transparency for radio pulsations when the pulsar is moving around the orbital phases 0.6 to 0.9.

 The transparency window arises for the following reasons. According to the proposed orbital geometry (Figure~\ref{fig:fitting-i60-orbit}), the pulsar appears in front of the Be disk as seen from the observer during the orbital phase of $\sim 0.5-0.9$, suggesting that the Be disk does not contribute to the DM within this phase interval. Additionally, Figure~\ref{fig:nh} is calculated with the system inclination angle of $\theta_{\rm o}=\pi/3$, the observer direction of $\nu_{\rm o}=1.29$~radian, and the momentum ratio of $\eta=0.18$ (Table~2). With these parameters, we find that the line of sight from the pulsar during the phase interval of $\sim 0.6-0.9$ is within the shock-cone and it does not pass through the stellar wind region.

 \cite{2012Caellas} investigate free-free absorption of the radio  emission from \lsi61 and estimate the pulsed flux density across different orbital phases and frequencies. Their calculation, based on the orbital parameter determined by
   \cite{2005casares}, predicts that  the  free-free absorption is strongest near periastron (orbital phase $\sim 0.23$). They also
   suggest that the radio pulses might be detectable around orbital phases 0.6-0.7, where the pulsar is located far distance  from the companion star. Similarly, \cite{2011McSwainRadio}  examine the free-free absorption  in the S-bands (1650-2250~MHz), C-bands (4400-5200 MHz) and X-bands~(8500-9300~MHz), and conclude that in the S and C bands, free-free absorption is too strong to allow  the pulsar detection at any orbital phase.  Only the X-bands provides an opportunity for an optically thin window  of the pulsar during the orbital phase of $0.4<\phi< 0.8$.  We note that while the detectable phases in these previous  studies are similar to our results,  the reason  of the radio escape is different process. In \cite{2012Caellas} and  \cite{2011McSwainRadio}, based on the orbital parameter determined by \cite{2005casares},   the detectability is related to the pulsar being at a large separation from the companion star.
   In our orbital parameter, the radio emission is detectable near the periastron, where the pulsar wind blocks the stellar wind traveling toward the observer.

 The proposed orbital geometry with the transparency window might align  with the FAST detection of the pulsed radio emission at near orbital phase of $\sim 0.6$. However, the transient nature  of the radio signal remains unexplained in the current study. In addition, several theoretical uncertainties exist. First, we find that the current polarization study cannot constrain the inclination angle $\nu_{\rm o}$, as mentioned in section~\ref{sec:lsi61}.
 If $\nu_{\rm o}$ is sufficiently small, the line of the sight lies entirely outside the shock-cone throughout the orbital and the no transparency window exists. Second, the previous study~\citep[e.g.][]{2013A&A...551A..17Z} suggests that
 the orbital motion generate a spiral structure of the shocked wind at the distance beyond $r>(2-3)a$ from the companion star. This spiral material has been neglected in the current study, but could introduce an additional DM, which is main contributor at the transparency window in Figure~\ref{fig:fitting-i60-orbit}. Consequently, a further multi-wavelength study dealing optical, X-ray and gamma-ray emission modeling, simultaneously, and a 3-D study for the stellar wind and pulsar wind interaction will be necessary to lead any conclusion for the orbital geometry as well as the origin of the transient pulsed emission from \lsi61.

{Finally, the non-thermal emissions from \lsi61 exhibit  super-orbital modulation with a period of $\sim 4.6$~years~\citep{1989ApJ...339.1054G, 2012ApJ...747L..29C,2013ApJ...773L..35A,2016A&A...591A..76A,2023MNRAS.525.2202C}. 
   It  has been modeled that  the super-orbital modulation  attributes to the precession of the Be disk \citep{2024ApJ...973..162C} or of the jet~\citep{2024A&A...683A.228J}.  Within the framework of our current study, if the Be disk undergoes precession, we expect that it would 
   induce a super-orbital modulation of polarization. In particular, if the precession angle is large, it would significantly affect to both baseline polarization degree and angle in Figure~\ref{fig:fitting-i60}.  We have analyzed  the polarization data reported by \cite{2020A&A...643A.170K}, which spans nearly continuous three years. This observation duration covers significant portion of the super-orbital period. If the super-orbital modulation is indeed driven by the Be disk precession, a more detailed examination of the polarization data could reveal a key parameters of the precession geometry.  A investigation involving Thomson scattering in a precessing Be disk will be presented in subsequent studies. 

 \section{Summary}
\label{sec:summary}
We have discussed the optical polarization of the gamma-ray binary system within the framework of the pulsar binary scenario, in which the pulsar wind and stellar wind create the shock. The optical linear polarization attributes to the Thomson scattering of the stellar photons by stellar wind and/or Be disk.  In this study, we developed the method for the calculation of the Thomson scattering of the stellar wind that accounts for the finite size of the companion star. Our analysis revealed that the influence of finite size would be significant for a compact system such like LS~5039, but would be negligible for a wider binary system such as \lsi61. We constrained the possible system parameters of \lsi61 by fitting the observed optical polarization using our model, and suggested the eccentricity of $e\sim 0.1$, which is consistent with the value derived by \citep{2020A&A...643A.170K}. The fitting result also suggested the periastron phase of $\nu_{\rm p} \sim 0.54$, which is closer to the phase where FAST detected the pulsed signal compared to previous results. Additionally, the fitting suggested that the momentum ratio ($\eta$) of the pulsar wind to stellar wind is $\eta>0.1$ and the mass-loss rate of the companion star wind is $\dot{M_{\rm w}}\sim 2\times 10^{-6}{\rm M_\odot}~{\rm year^{-1}}$. Then we calculate that the magnetic field is $\sim1.5\times10^{14}\mathrm{G}$, which supports the scenario that the compact object in \lsi61 is a  highly-B pulsar or magnetar. We discussed the DM along the line of sight from the pulsar caused by the stellar wind and the Be disk and typically obtained the DM$\sim10^4~{\rm pc~cm^{-3}}$, which is substantially larger than the value measured during the 2020 FAST observation. On the other hand, we also found that the specific inclination angle of the system suggests a potential window of transparency for radio pulsations when the pulsar is moving around the orbital phases of 0.6 to 0.9. This would align with the orbital phase ($\sim0.6$) where FAST detected transient pulsed radio emissions, although further investigation of the 3-D wind structure is necessary. 

\begin{acknowledgements}
We acknowledge Dr. V.Kravtsov for providing us with the optical polarization data of \lsi61. We thank Dr A.M. Chen for useful discussions on the gamma-ray binary system. 
This work is supported by the National Key R\&D Program of China (Nos. 2020YFC2201400, SQ2023YFC220007) and the National Natural Science Foundation of China under grant 12473012. W.H.Lei. acknowledges support from science research grants from the China Manned Space Project with NO.CMS-CSST-2021-B11.
\end{acknowledgements}

\bibliography{references}

\begin{thebibliography}{}
\expandafter\ifx\csname natexlab\endcsname\relax\def\natexlab#1{#1}\fi
\providecommand{\url}[1]{\href{#1}{#1}}
\providecommand{\dodoi}[1]{doi:~\href{http://doi.org/#1}{\nolinkurl{#1}}}
\providecommand{\doeprint}[1]{\href{http://ascl.net/#1}{\nolinkurl{http://ascl.net/#1}}}
\providecommand{\doarXiv}[1]{\href{https://arxiv.org/abs/#1}{\nolinkurl{https://arxiv.org/abs/#1}}}

\bibitem[{{Abeysekara} {et~al.}(2018{\natexlab{a}}){Abeysekara}, {Benbow},
  {Bird}, {Brill}, {Brose}, {Buckley}, {Chromey}, {Daniel}, {Falcone},
  {Finley}, {Fortson}, {Furniss}, {Gent}, {Gillanders}, {Hanna}, {Hassan},
  {Hervet}, {Holder}, {Hughes}, {Humensky}, {Kaaret}, {Kar}, {Kertzman},
  {Kieda}, {Krause}, {Krennrich}, {Kumar}, {Lang}, {Lin}, {Maier}, {Moriarty},
  {Mukherjee}, {O'Brien}, {Ong}, {Otte}, {Park}, {Petrashyk}, {Pohl},
  {Pueschel}, {Quinn}, {Ragan}, {Richards}, {Roache}, {Sadeh}, {Santander},
  {Schlenstedt}, {Sembroski}, {Sushch}, {Tyler}, {Vassiliev}, {Wakely},
  {Weinstein}, {Wells}, {Wilcox}, {Wilhelm}, {Williams}, {Williamson},
  {Zitzer}, {VERITAS Collaboration}, {Acciari}, {Ansoldi}, {Antonelli}, {Arbet
  Engels}, {Baack}, {Babi{\'c}}, {Banerjee}, {Barres de Almeida}, {Barrio},
  {Becerra Gonz{\'a}lez}, {Bednarek}, {Bernardini}, {Berti}, {Besenrieder},
  {Bhattacharyya}, {Bigongiari}, {Biland}, {Blanch}, {Bonnoli}, {Busetto},
  {Carosi}, {Ceribella}, {Cikota}, {Colak}, {Colin}, {Colombo}, {Contreras},
  {Cortina}, {Covino}, {D'Elia}, {Da Vela}, {Dazzi}, {De Angelis}, {De Lotto},
  {Delfino}, {Delgado}, {Di Pierro}, {Do Souto Espi{\~n}era}, {Dom{\'\i}nguez},
  {Dominis Prester}, {Dorner}, {Doro}, {Einecke}, {Elsaesser}, {Fallah
  Ramazani}, {Fattorini}, {Fern{\'a}ndez-Barral}, {Ferrara}, {Fidalgo},
  {Foffano}, {Fonseca}, {Font}, {Fruck}, {Galindo}, {Gallozzi}, {Garc{\'\i}a
  L{\'o}pez}, {Garczarczyk}, {Gasparyan}, {Gaug}, {Giammaria}, {Godinovi{\'c}},
  {Guberman}, {Hadasch}, {Hahn}, {Herrera}, {Hoang}, {Hrupec}, {Inoue},
  {Ishio}, {Iwamura}, {Kubo}, {Kushida}, {Kuve{\v{z}}di{\'c}}, {Lamastra},
  {Lelas}, {Leone}, {Lindfors}, {Lombardi}, {Longo}, {L{\'o}pez},
  {L{\'o}pez-Oramas}, {Machado de Oliveira Fraga}, {Maggio}, {Majumdar},
  {Makariev}, {Mallamaci}, {Maneva}, {Manganaro}, {Mannheim}, {Maraschi},
  {Mariotti}, {Mart{\'\i}nez}, {Masuda}, {Mazin}, {Minev}, {Miranda},
  {Mirzoyan}, {Molina}, {Moralejo}, {Moreno}, {Moretti}, {Munar-Adrover},
  {Neustroev}, {Niedzwiecki}, {Nievas Rosillo}, {Nigro}, {Nilsson}, {Ninci},
  {Nishijima}, {Noda}, {Nogu{\'e}s}, {N{\"o}the}, {Paiano}, {Palacio},
  {Paneque}, {Paoletti}, {Paredes}, {Pedaletti}, {Pe{\~n}il}, {Peresano},
  {Persic}, {Prada Moroni}, {Prandini}, {Puljak}, {Garcia}, {Rhode},
  {Rib{\'o}}, {Rico}, {Righi}, {Rugliancich}, {Saha}, {Sahakyan}, {Saito},
  {Satalecka}, {Schweizer}, {Sitarek}, {{\v{S}}nidari{\'c}}, {Sobczynska},
  {Somero}, {Stamerra}, \& {Strzys}}]{2018ApJ...867L..19A}
{Abeysekara}, A.~U., {Benbow}, W., {Bird}, R., {et~al.} 2018{\natexlab{a}},
  \apjl, 867, L19, \dodoi{10.3847/2041-8213/aae70e}

\bibitem[{{Abeysekara} {et~al.}(2018{\natexlab{b}}){Abeysekara}, {Albert},
  {Alfaro}, {Alvarez}, {{\'A}lvarez}, {Arceo}, {Arteaga-Vel{\'a}zquez}, {Avila
  Rojas}, {Ayala Solares}, {Belmont-Moreno}, {BenZvi}, {Brisbois},
  {Caballero-Mora}, {Capistr{\'a}n}, {Carrami{\~n}ana}, {Casanova}, {Castillo},
  {Cotti}, {Cotzomi}, {Couti{\~n}o de Le{\'o}n}, {De Le{\'o}n}, {De la Fuente},
  {D{\'\i}az-V{\'e}lez}, {Dichiara}, {Dingus}, {DuVernois}, {Ellsworth},
  {Engel}, {Espinoza}, {Fang}, {Fleischhack}, {Fraija}, {Galv{\'a}n-G{\'a}mez},
  {Garc{\'\i}a-Gonz{\'a}lez}, {Garfias}, {Gonz{\'a}lez-Mu{\~n}oz},
  {Gonz{\'a}lez}, {Goodman}, {Hampel-Arias}, {Harding}, {Hernandez}, {Hinton},
  {Hona}, {Hueyotl-Zahuantitla}, {Hui}, {H{\"u}ntemeyer}, {Iriarte},
  {Jardin-Blicq}, {Joshi}, {Kaufmann}, {Kar}, {Kunde}, {Lauer}, {Lee},
  {Le{\'o}n Vargas}, {Li}, {Linnemann}, {Longinotti}, {Luis-Raya},
  {L{\'o}pez-Coto}, {Malone}, {Marinelli}, {Martinez}, {Martinez-Castellanos},
  {Mart{\'\i}nez-Castro}, {Matthews}, {Miranda-Romagnoli}, {Moreno},
  {Mostaf{\'a}}, {Nayerhoda}, {Nellen}, {Newbold}, {Nisa}, {Noriega-Papaqui},
  {Pretz}, {P{\'e}rez-P{\'e}rez}, {Ren}, {Rho}, {Rivi{\`e}re},
  {Rosa-Gonz{\'a}lez}, {Rosenberg}, {Ruiz-Velasco}, {Salesa Greus}, {Sandoval},
  {Schneider}, {Schoorlemmer}, {Seglar Arroyo}, {Sinnis}, {Smith}, {Springer},
  {Surajbali}, {Taboada}, {Tibolla}, {Tollefson}, {Torres}, {Vianello},
  {Villase{\~n}or}, {Weisgarber}, {Werner}, {Westerhoff}, {Wood}, {Yapici},
  {Yodh}, {Zepeda}, {Zhang}, \& {Zhou}}]{2018Natur.562...82A}
{Abeysekara}, A.~U., {Albert}, A., {Alfaro}, R., {et~al.} 2018{\natexlab{b}},
  \nat, 562, 82, \dodoi{10.1038/s41586-018-0565-5}

\bibitem[{{Ackermann} {et~al.}(2013){Ackermann}, {Ajello}, {Ballet},
  {Barbiellini}, {Bastieri}, {Bellazzini}, {Bonamente}, {Brandt}, {Bregeon},
  {Brigida}, {Bruel}, {Buehler}, {Buson}, {Caliandro}, {Cameron}, {Caraveo},
  {Casandjian}, {Cavazzuti}, {Cecchi}, {Chekhtman}, {Chiang}, {Chiaro},
  {Ciprini}, {Claus}, {Cohen-Tanugi}, {Cominsky}, {Conrad}, {Cutini}, {Dalton},
  {D'Ammando}, {de Angelis}, {den Hartog}, {de Palma}, {Dermer}, {Digel}, {Di
  Venere}, {Drell}, {Dubois}, {Favuzzi}, {Fegan}, {Ferrara}, {Focke},
  {Franckowiak}, {Funk}, {Fusco}, {Gargano}, {Gasparrini}, {Germani},
  {Giglietto}, {Giordano}, {Giroletti}, {Glanzman}, {Godfrey}, {Grenier},
  {Guiriec}, {Hadasch}, {Hanabata}, {Harding}, {Hayashida}, {Hays}, {Hill},
  {Horan}, {Hughes}, {Jogler}, {J{\'o}hannesson}, {Johnson}, {Johnson},
  {Kawano}, {Kerr}, {Kn{\"o}dlseder}, {Kuss}, {Lande}, {Larsson}, {Latronico},
  {Lemoine-Goumard}, {Li}, {Longo}, {Lovellette}, {Lubrano}, {Mayer},
  {Mazziotta}, {McEnery}, {Michelson}, {Mizuno}, {Monzani}, {Morselli},
  {Moskalenko}, {Murgia}, {Nemmen}, {Nuss}, {Ohsugi}, {Okumura}, {Orienti},
  {Orlando}, {Ormes}, {Paneque}, {Papitto}, {Perkins}, {Pesce-Rollins},
  {Piron}, {Pivato}, {Rain{\`o}}, {Rando}, {Razzano}, {Rea}, {Reimer},
  {Reimer}, {Scargle}, {Schulz}, {Sgr{\`o}}, {Siskind}, {Spandre}, {Spinelli},
  {Takahashi}, {Thayer}, {Thayer}, {Tinivella}, {Torres}, {Tosti}, {Troja},
  {Uchiyama}, {Usher}, {Vandenbroucke}, {Vasileiou}, {Vianello}, {Vitale},
  {Werner}, {Winer}, \& {Wood}}]{2013ApJ...773L..35A}
{Ackermann}, M., {Ajello}, M., {Ballet}, J., {et~al.} 2013, \apjl, 773, L35,
  \dodoi{10.1088/2041-8205/773/2/L35}

\bibitem[{{Aharonian} {et~al.}(2005){Aharonian}, {Akhperjanian}, {Aye},
  {Bazer-Bachi}, {Beilicke}, {Benbow}, {Berge}, {Berghaus}, {Bernl{\"o}hr},
  {Boisson}, {Bolz}, {Braun}, {Breitling}, {Brown}, {Bussons Gordo},
  {Chadwick}, {Chounet}, {Cornils}, {Costamante}, {Degrange},
  {Djannati-Ata{\"\i}}, {O'C. Drury}, {Dubus}, {Emmanoulopoulos}, {Espigat},
  {Feinstein}, {Fleury}, {Fontaine}, {Fuchs}, {Funk}, {Gallant}, {Giebels},
  {Gillessen}, {Glicenstein}, {Goret}, {Hadjichristidis}, {Hauser},
  {Heinzelmann}, {Henri}, {Hermann}, {Hinton}, {Hofmann}, {Holleran}, {Horns},
  {de Jager}, {Johnston}, {Kh{\'e}lifi}, {Kirk}, {Komin}, {Konopelko},
  {Latham}, {Le Gallou}, {Lemi{\`e}re}, {Lemoine-Goumard}, {Leroy},
  {Martineau-Huynh}, {Lohse}, {Marcowith}, {Masterson}, {McComb}, {de Naurois},
  {Nolan}, {Noutsos}, {Orford}, {Osborne}, {Ouchrif}, {Panter}, {Pelletier},
  {Pita}, {P{\"u}hlhofer}, {Punch}, {Raubenheimer}, {Raue}, {Raux}, {Rayner},
  {Redondo}, {Reimer}, {Reimer}, {Ripken}, {Rob}, {Rolland}, {Rowell},
  {Sahakian}, {Saug{\'e}}, {Schlenker}, {Schlickeiser}, {Schuster}, {Schwanke},
  {Siewert}, {Skj{\ae}raasen}, {Sol}, {Steenkamp}, {Stegmann}, {Tavernet},
  {Terrier}, {Th{\'e}oret}, {Tluczykont}, {Vasileiadis}, {Venter}, {Vincent},
  {V{\"o}lk}, \& {Wagner}}]{2005A&A...442....1A}
{Aharonian}, F., {Akhperjanian}, A.~G., {Aye}, K.~M., {et~al.} 2005, \aap, 442,
  1, \dodoi{10.1051/0004-6361:20052983}

\bibitem[{{Ahnen} {et~al.}(2016){Ahnen}, {Ansoldi}, {Antonelli}, {Antoranz},
  {Babic}, {Banerjee}, {Bangale}, {Barres de Almeida}, {Barrio}, {Becerra
  Gonz{\'a}lez}, {Bednarek}, {Bernardini}, {Biasuzzi}, {Biland}, {Blanch},
  {Bonnefoy}, {Bonnoli}, {Borracci}, {Bretz}, {Buson}, {Carosi}, {Chatterjee},
  {Clavero}, {Colin}, {Colombo}, {Contreras}, {Cortina}, {Covino}, {Da Vela},
  {Dazzi}, {De Angelis}, {De Lotto}, {de O{\~n}a Wilhelmi}, {Delgado Mendez},
  {Di Pierro}, {Dom{\'\i}nguez}, {Dominis Prester}, {Dorner}, {Doro},
  {Einecke}, {Eisenacher Glawion}, {Elsaesser}, {Fern{\'a}ndez-Barral},
  {Fidalgo}, {Fonseca}, {Font}, {Frantzen}, {Fruck}, {Galindo}, {Garc{\'\i}a
  L{\'o}pez}, {Garczarczyk}, {Garrido Terrats}, {Gaug}, {Giammaria},
  {Godinovi{\'c}}, {Gonz{\'a}lez Mu{\~n}oz}, {Gora}, {Guberman}, {Hadasch},
  {Hahn}, {Hanabata}, {Hayashida}, {Herrera}, {Hose}, {Hrupec}, {Hughes},
  {Idec}, {Kodani}, {Konno}, {Kubo}, {Kushida}, {La Barbera}, {Lelas},
  {Lindfors}, {Lombardi}, {Longo}, {L{\'o}pez}, {L{\'o}pez-Coto},
  {L{\'o}pez-Oramas}, {Majumdar}, {Makariev}, {Mallot}, {Maneva}, {Manganaro},
  {Mannheim}, {Maraschi}, {Marcote}, {Mariotti}, {Mart{\'\i}nez}, {Mazin},
  {Menzel}, {Miranda}, {Mirzoyan}, {Moralejo}, {Moretti}, {Nakajima},
  {Neustroev}, {Niedzwiecki}, {Nievas Rosillo}, {Nilsson}, {Nishijima}, {Noda},
  {Orito}, {Overkemping}, {Paiano}, {Palacio}, {Palatiello}, {Paneque},
  {Paoletti}, {Paredes}, {Paredes-Fortuny}, {Pedaletti}, {Persic}, {Poutanen},
  {Prada Moroni}, {Prandini}, {Puljak}, {Rhode}, {Rib{\'o}}, {Rico}, {Rodriguez
  Garcia}, {Saito}, {Satalecka}, {Schultz}, {Schweizer}, {Shore},
  {Sillanp{\"a}{\"a}}, {Sitarek}, {Snidaric}, {Sobczynska}, {Stamerra},
  {Steinbring}, {Strzys}, {Takalo}, {Takami}, {Tavecchio}, {Temnikov},
  {Terzi{\'c}}, {Tescaro}, {Teshima}, {Thaele}, {Torres}, {Toyama}, {Treves},
  {Verguilov}, {Vovk}, {Ward}, {Will}, {Wu}, {Zanin}, {MAGIC Collaboration},
  {Casares}, \& {Herrero}}]{2016A&A...591A..76A}
{Ahnen}, M.~L., {Ansoldi}, S., {Antonelli}, L.~A., {et~al.} 2016, \aap, 591,
  A76, \dodoi{10.1051/0004-6361/201527964}

\bibitem[{{Albert} {et~al.}(2007){Albert}, {Aliu}, {Anderhub}, {Antoranz},
  {Armada}, {Baixeras}, {Barrio}, {Bartko}, {Bastieri}, {Becker}, {Bednarek},
  {Berger}, {Bigongiari}, {Biland}, {Bock}, {Bordas}, {Bosch-Ramon}, {Bretz},
  {Britvitch}, {Camara}, {Carmona}, {Chilingarian}, {Coarasa}, {Commichau},
  {Contreras}, {Cortina}, {Costado}, {Curtef}, {Danielyan}, {Dazzi}, {De
  Angelis}, {Delgado}, {de los Reyes}, {De Lotto}, {Domingo-Santamar{\'\i}a},
  {Dorner}, {Doro}, {Errando}, {Fagiolini}, {Ferenc}, {Fern{\'a}ndez}, {Firpo},
  {Flix}, {Fonseca}, {Font}, {Fuchs}, {Galante}, {Garc{\'\i}a-L{\'o}pez},
  {Garczarczyk}, {Gaug}, {Giller}, {Goebel}, {Hakobyan}, {Hayashida},
  {Hengstebeck}, {Herrero}, {H{\"o}hne}, {Hose}, {Hsu}, {Jacon}, {Jogler},
  {Kosyra}, {Kranich}, {Kritzer}, {Laille}, {Lindfors}, {Lombardi}, {Longo},
  {L{\'o}pez}, {L{\'o}pez}, {Lorenz}, {Majumdar}, {Maneva}, {Mannheim},
  {Mansutti}, {Mariotti}, {Mart{\'\i}nez}, {Mazin}, {Merck}, {Meucci}, {Meyer},
  {Miranda}, {Mirzoyan}, {Mizobuchi}, {Moralejo}, {Nieto}, {Nilsson},
  {Ninkovic}, {O{\~n}a-Wilhelmi}, {Otte}, {Oya}, {Panniello}, {Paoletti},
  {Paredes}, {Pasanen}, {Pascoli}, {Pauss}, {Pegna}, {Persic}, {Peruzzo},
  {Piccioli}, {Prandini}, {Puchades}, {Raymers}, {Rhode}, {Rib{\'o}}, {Rico},
  {Rissi}, {Robert}, {R{\"u}gamer}, {Saggion}, {Saito}, {S{\'a}nchez},
  {Sartori}, {Scalzotto}, {Scapin}, {Schmitt}, {Schweizer}, {Shayduk},
  {Shinozaki}, {Shore}, {Sidro}, {Sillanp{\"a}{\"a}}, {Sobczynska}, {Stamerra},
  {Stark}, {Takalo}, {Temnikov}, {Tescaro}, {Teshima}, {Torres}, {Turini},
  {Vankov}, {Vitale}, {Wagner}, {Wibig}, {Wittek}, {Zandanel}, {Zanin}, \&
  {Zapatero}}]{2007ApJ...665L..51A}
{Albert}, J., {Aliu}, E., {Anderhub}, H., {et~al.} 2007, \apjl, 665, L51,
  \dodoi{10.1086/521145}

\bibitem[{{Alfaro} {et~al.}(2024){Alfaro}, {Alvarez}, {Arteaga-Vel{\'a}zquez},
  {Avila Rojas}, {Ayala Solares}, {Babu}, {Belmont-Moreno}, {Caballero-Mora},
  {Capistr{\'a}n}, {Carrami{\~n}ana}, {Casanova}, {Cotti}, {Cotzomi},
  {Couti{\~n}o de Le{\'o}n}, {De la Fuente}, {Depaoli}, {Di Lalla}, {Diaz
  Hernandez}, {Dingus}, {DuVernois}, {Durocher}, {D{\'\i}az-V{\'e}lez},
  {Engel}, {Espinoza}, {Fan}, {Fang}, {Fraija}, {Fraija},
  {Garc{\'\i}a-Gonz{\'a}lez}, {Garfias}, {Gonzalez Mu{\~n}oz}, {Gonz{\'a}lez},
  {Goodman}, {Groetsch}, {Harding}, {Herzog}, {Hinton}, {Huang},
  {Hueyotl-Zahuantitla}, {H{\"u}ntemeyer}, {Iriarte}, {Joshi}, {Kaufmann},
  {Kieda}, {de Le{\'o}n}, {Lee}, {Le{\'o}n Vargas}, {Linnemann}, {Longinotti},
  {Luis-Raya}, {Malone}, {Martinez}, {Mart{\'\i}nez-Castro}, {Matthews},
  {Miranda-Romagnoli}, {Morales-Soto}, {Moreno}, {Mostaf{\'a}}, {Nayerhoda},
  {Nellen}, {Newbold}, {Nisa}, {Noriega-Papaqui}, {Olivera-Nieto}, {Omodei},
  {Osorio}, {P{\'e}rez Araujo}, {P{\'e}rez-P{\'e}rez}, {Rho},
  {Rosa-Gonz{\'a}lez}, {Ruiz-Velasco}, {Salazar}, {Salazar-Gallegos},
  {Sandoval}, {Schneider}, {Serna-Franco}, {Smith}, {Son}, {Springer},
  {Tibolla}, {Tollefson}, {Torres}, {Torres-Escobedo}, {Turner},
  {Ure{\~n}a-Mena}, {Varela}, {Villase{\~n}or}, {Wang}, {Watson}, {Willox},
  {Yun-C{\'a}rcamo}, \& {Zhou}}]{2024Natur.634..557A}
{Alfaro}, R., {Alvarez}, C., {Arteaga-Vel{\'a}zquez}, J.~C., {et~al.} 2024,
  \nat, 634, 557, \dodoi{10.1038/s41586-024-07995-9}

\bibitem[{{Anderhub} {et~al.}(2009){Anderhub}, {Antonelli}, {Antoranz},
  {Backes}, {Baixeras}, {Balestra}, {Barrio}, {Bastieri}, {Becerra
  Gonz{\'a}lez}, {Becker}, {Bednarek}, {Berger}, {Bernardini}, {Biland},
  {Blanch Bigas}, {Bock}, {Bonnoli}, {Bordas}, {Borla Tridon}, {Bosch-Ramon},
  {Bose}, {Braun}, {Bretz}, {Britzger}, {Camara}, {Carmona}, {Carosi}, {Colin},
  {Commichau}, {Contreras}, {Cortina}, {Costado}, {Covino}, {Dazzi}, {De
  Angelis}, {de Cea del Pozo}, {De los Reyes}, {De Lotto}, {De Maria}, {De
  Sabata}, {Delgado Mendez}, {Dom{\'\i}nguez}, {Dominis Prester}, {Dorner},
  {Doro}, {Elsaesser}, {Errando}, {Ferenc}, {Fern{\'a}ndez}, {Firpo},
  {Fonseca}, {Font}, {Galante}, {Garc{\'\i}a L{\'o}pez}, {Garczarczyk}, {Gaug},
  {Godinovic}, {Goebel}, {Hadasch}, {Herrero}, {Hildebrand},
  {H{\"o}hne-M{\"o}nch}, {Hose}, {Hrupec}, {Hsu}, {Jogler}, {Klepser},
  {Kranich}, {La Barbera}, {Laille}, {Leonardo}, {Lindfors}, {Lombardi},
  {Longo}, {L{\'o}pez}, {Lorenz}, {Majumdar}, {Maneva}, {Mankuzhiyil},
  {Mannheim}, {Maraschi}, {Mariotti}, {Mart{\'\i}nez}, {Mazin}, {Meucci},
  {Miranda}, {Mirzoyan}, {Miyamoto}, {Mold{\'o}n}, {Moles}, {Moralejo},
  {Nieto}, {Nilsson}, {Ninkovic}, {Orito}, {Oya}, {Paoletti}, {Paredes},
  {Pasanen}, {Pascoli}, {Pauss}, {Pegna}, {Perez-Torres}, {Persic}, {Peruzzo},
  {Prada}, {Prandini}, {Puchades}, {Puljak}, {Reichardt}, {Rhode}, {Rib{\'o}},
  {Rico}, {Rissi}, {Robert}, {R{\"u}gamer}, {Saggion}, {Saito}, {Salvati},
  {S{\'a}nchez-Conde}, {Satalecka}, {Scalzotto}, {Scapin}, {Schweizer},
  {Shayduk}, {Shore}, {Sidro}, {Sierpowska-Bartosik}, {Sillanp{\"a}{\"a}},
  {Sitarek}, {Sobczynska}, {Spanier}, {Spiro}, {Stamerra}, {Stark}, {Suric},
  {Takalo}, {Tavecchio}, {Temnikov}, {Tescaro}, {Teshima}, {Torres}, {Turini},
  {Vankov}, {Wagner}, {Zabalza}, {Zandanel}, {Zanin}, {Zapatero}, {MAGIC
  Collaboration}, {Falcone}, {Vetere}, {Gehrels}, {Trushkin}, {Dhawan}, \&
  {Reig}}]{2009ApJ...706L..27A}
{Anderhub}, H., {Antonelli}, L.~A., {Antoranz}, P., {et~al.} 2009, \apjl, 706,
  L27, \dodoi{10.1088/0004-637X/706/1/L27}

\bibitem[{{Aragona} {et~al.}(2009){Aragona}, {McSwain}, {Grundstrom}, {Marsh},
  {Roettenbacher}, {Hessler}, {Boyajian}, \& {Ray}}]{2009ApJ...698..514A}
{Aragona}, C., {McSwain}, M.~V., {Grundstrom}, E.~D., {et~al.} 2009, \apj, 698,
  514, \dodoi{10.1088/0004-637X/698/1/514}

\bibitem[{{Archibald} {et~al.}(2016){Archibald}, {Kaspi}, {Tendulkar}, \&
  {Scholz}}]{2016ApJ...829L..21A}
{Archibald}, R.~F., {Kaspi}, V.~M., {Tendulkar}, S.~P., \& {Scholz}, P. 2016,
  \apjl, 829, L21, \dodoi{10.3847/2041-8205/829/1/L21}

\bibitem[{{Bosch-Ramon}(2013)}]{2013A&A...560A..32B}
{Bosch-Ramon}, V. 2013, \aap, 560, A32, \dodoi{10.1051/0004-6361/201322249}

\bibitem[{{Bosch-Ramon} {et~al.}(2012){Bosch-Ramon}, {Barkov}, {Khangulyan}, \&
  {Perucho}}]{2012A&A...544A..59B}
{Bosch-Ramon}, V., {Barkov}, M.~V., {Khangulyan}, D., \& {Perucho}, M. 2012,
  \aap, 544, A59, \dodoi{10.1051/0004-6361/201219251}

\bibitem[{{Bosch-Ramon} \& {Paredes}(2004)}]{2004A&A...425.1069B}
{Bosch-Ramon}, V., \& {Paredes}, J.~M. 2004, \aap, 425, 1069,
  \dodoi{10.1051/0004-6361:20041185}

\bibitem[{{Brown} {et~al.}(1978){Brown}, {McLean}, \&
  {Emslie}}]{1978A&A....68..415B}
{Brown}, J.~C., {McLean}, I.~S., \& {Emslie}, A.~G. 1978, \aap, 68, 415

\bibitem[{{Ca{\~n}ellas} {et~al.}(2012){Ca{\~n}ellas}, {Joshi}, {Paredes},
  {Ishwara-Chandra}, {Mold{\'o}n}, {Zabalza}, {Mart{\'\i}}, \&
  {Rib{\'o}}}]{2012Caellas}
{Ca{\~n}ellas}, A., {Joshi}, B.~C., {Paredes}, J.~M., {et~al.} 2012, \aap, 543,
  A122, \dodoi{10.1051/0004-6361/201117619}

\bibitem[{{Cant{\'o}} {et~al.}(1996){Cant{\'o}}, {Raga}, \&
  {Wilkin}}]{1996ApJ...469..729C}
{Cant{\'o}}, J., {Raga}, A.~C., \& {Wilkin}, F.~P. 1996, \apj, 469, 729,
  \dodoi{10.1086/177820}

\bibitem[{{Carciofi} \& {Bjorkman}(2008)}]{2008ApJ...684.1374C}
{Carciofi}, A.~C., \& {Bjorkman}, J.~E. 2008, \apj, 684, 1374,
  \dodoi{10.1086/589875}

\bibitem[{{Casares} {et~al.}(2005){Casares}, {Ribas}, {Paredes}, {Mart{\'\i}},
  \& {Allende Prieto}}]{2005MNRAS.360.1105C}
{Casares}, J., {Ribas}, I., {Paredes}, J.~M., {Mart{\'\i}}, J., \& {Allende
  Prieto}, C. 2005, \mnras, 360, 1105, \dodoi{10.1111/j.1365-2966.2005.09106.x}

\bibitem[{Casares {et~al.}(2005)Casares, Rib{\'o}, Ribas, Paredes, Mart{\'\i},
  \& Herrero}]{2005casares}
Casares, J., Rib{\'o}, M., Ribas, I., {et~al.} 2005, Monthly Notices of the
  Royal Astronomical Society, 364, 899

\bibitem[{{Cerutti} {et~al.}(2008){Cerutti}, {Dubus}, \&
  {Henri}}]{2008A&A...488...37C}
{Cerutti}, B., {Dubus}, G., \& {Henri}, G. 2008, \aap, 488, 37,
  \dodoi{10.1051/0004-6361:200809939}

\bibitem[{{Chen} {et~al.}(2021){Chen}, {Guo}, {Yu}, \&
  {Takata}}]{2021A&A...652A..39C}
{Chen}, A.~M., {Guo}, Y.~D., {Yu}, Y.~W., \& {Takata}, J. 2021, \aap, 652, A39,
  \dodoi{10.1051/0004-6361/202140951}

\bibitem[{{Chen} \& {Takata}(2022)}]{2022A&A...658A.153C}
{Chen}, A.~M., \& {Takata}, J. 2022, \aap, 658, A153,
  \dodoi{10.1051/0004-6361/202142258}

\bibitem[{{Chen} {et~al.}(2024){Chen}, {Takata}, \& {Yu}}]{2024ApJ...973..162C}
{Chen}, A.~M., {Takata}, J., \& {Yu}, Y.~W. 2024, \apj, 973, 162,
  \dodoi{10.3847/1538-4357/ad6b0a}

\bibitem[{{Chernyakova} {et~al.}(2017){Chernyakova}, {Babyk}, {Malyshev},
  {Vovk}, {Tsygankov}, {Takahashi}, \& {Fukazawa}}]{2017MNRAS.470.1718C}
{Chernyakova}, M., {Babyk}, I., {Malyshev}, D., {et~al.} 2017, \mnras, 470,
  1718, \dodoi{10.1093/mnras/stx1335}

\bibitem[{{Chernyakova} {et~al.}(2023){Chernyakova}, {Malyshev}, {Neronov}, \&
  {Savchenko}}]{2023MNRAS.525.2202C}
{Chernyakova}, M., {Malyshev}, D., {Neronov}, A., \& {Savchenko}, D. 2023,
  \mnras, 525, 2202, \dodoi{10.1093/mnras/stad2380}

\bibitem[{{Chernyakova} {et~al.}(2012){Chernyakova}, {Neronov}, {Molkov},
  {Malyshev}, {Lutovinov}, \& {Pooley}}]{2012ApJ...747L..29C}
{Chernyakova}, M., {Neronov}, A., {Molkov}, S., {et~al.} 2012, \apjl, 747, L29,
  \dodoi{10.1088/2041-8205/747/2/L29}

\bibitem[{{Chernyakova} {et~al.}(2015){Chernyakova}, {Neronov}, {van Soelen},
  {Callanan}, {O'Shaughnessy}, {Babyk}, {Tsygankov}, {Vovk}, {Krivonos},
  {Tomsick}, {Malyshev}, {Li}, {Wood}, {Torres}, {Zhang}, {Kretschmar},
  {McSwain}, {Buckley}, \& {Koen}}]{2015MNRAS.454.1358C}
{Chernyakova}, M., {Neronov}, A., {van Soelen}, B., {et~al.} 2015, \mnras, 454,
  1358, \dodoi{10.1093/mnras/stv1988}

\bibitem[{{Combi} {et~al.}(2004){Combi}, {Cellone}, {Mart{\'\i}}, {Rib{\'o}},
  {Mirabel}, \& {Casares}}]{2004A&A...427..959C}
{Combi}, J.~A., {Cellone}, S.~A., {Mart{\'\i}}, J., {et~al.} 2004, \aap, 427,
  959, \dodoi{10.1051/0004-6361:20041433}

\bibitem[{{Corbet} {et~al.}(2016){Corbet}, {Chomiuk}, {Coe}, {Coley}, {Dubus},
  {Edwards}, {Martin}, {McBride}, {Stevens}, {Strader}, {Townsend}, \&
  {Udalski}}]{2016ApJ...829..105C}
{Corbet}, R.~H.~D., {Chomiuk}, L., {Coe}, M.~J., {et~al.} 2016, \apj, 829, 105,
  \dodoi{10.3847/0004-637X/829/2/105}

\bibitem[{{Corbet} {et~al.}(2019){Corbet}, {Chomiuk}, {Coe}, {Coley}, {Dubus},
  {Edwards}, {Martin}, {McBride}, {Stevens}, {Strader}, \&
  {Townsend}}]{2019ApJ...884...93C}
---. 2019, \apj, 884, 93, \dodoi{10.3847/1538-4357/ab3e32}

\bibitem[{{Dubus}(2006)}]{2006A&A...456..801D}
{Dubus}, G. 2006, \aap, 456, 801, \dodoi{10.1051/0004-6361:20054779}

\bibitem[{{Dubus}(2013)}]{2013A&ARv..21...64D}
---. 2013, \aapr, 21, 64, \dodoi{10.1007/s00159-013-0064-5}

\bibitem[{{Dubus} {et~al.}(2010){Dubus}, {Cerutti}, \&
  {Henri}}]{2010A&A...516A..18D}
{Dubus}, G., {Cerutti}, B., \& {Henri}, G. 2010, \aap, 516, A18,
  \dodoi{10.1051/0004-6361/201014023}

\bibitem[{{Foreman-Mackey} {et~al.}(2013){Foreman-Mackey}, {Hogg}, {Lang}, \&
  {Goodman}}]{2013PASP..125..306F}
{Foreman-Mackey}, D., {Hogg}, D.~W., {Lang}, D., \& {Goodman}, J. 2013, \pasp,
  125, 306, \dodoi{10.1086/670067}

\bibitem[{{Gregory} \& {Taylor}(1978)}]{1978Natur.272..704G}
{Gregory}, P.~C., \& {Taylor}, A.~R. 1978, \nat, 272, 704,
  \dodoi{10.1038/272704a0}

\bibitem[{{Gregory} {et~al.}(1989){Gregory}, {Xu}, {Backhouse}, \&
  {Reid}}]{1989ApJ...339.1054G}
{Gregory}, P.~C., {Xu}, H.-J., {Backhouse}, C.~J., \& {Reid}, A. 1989, \apj,
  339, 1054, \dodoi{10.1086/167359}

\bibitem[{{Grundstrom} {et~al.}(2007){Grundstrom}, {Caballero-Nieves}, {Gies},
  {Huang}, {McSwain}, {Rafter}, {Riddle}, {Williams}, \&
  {Wingert}}]{2007ApJ...656..437G}
{Grundstrom}, E.~D., {Caballero-Nieves}, S.~M., {Gies}, D.~R., {et~al.} 2007,
  \apj, 656, 437, \dodoi{10.1086/510509}

\bibitem[{Halonen \& Jones(2013)}]{2013Halonen}
Halonen, R.~J., \& Jones, C.~E. 2013, The Astrophysical Journal, 765, 17,
  \dodoi{10.1088/0004-637X/765/1/17}

\bibitem[{{Halonen} {et~al.}(2013){Halonen}, {Mackay}, \&
  {Jones}}]{2013ApJS..204...11H}
{Halonen}, R.~J., {Mackay}, F.~E., \& {Jones}, C.~E. 2013, \apjs, 204, 11,
  \dodoi{10.1088/0067-0049/204/1/11}

\bibitem[{{HESS Collaboration} {et~al.}(2015){HESS Collaboration},
  {Abramowski}, {Acero}, {Aharonian}, {Ait Benkhali}, {Akhperjanian},
  {Ang{\"u}ner}, {Anton}, {Balenderan}, {Balzer}, {Barnacka}, {Becherini},
  {Becker Tjus}, {Bernl{\"o}hr}, {Birsin}, {Bissaldi}, {Biteau},
  {B{\"o}ttcher}, {Boisson}, {Bolmont}, {Bordas}, {Brucker}, {Brun}, {Brun},
  {Bulik}, {Carrigan}, {Casanova}, {Cerruti}, {Chadwick}, {Chalme-Calvet},
  {Chaves}, {Cheesebrough}, {Chr{\'e}tien}, {Clapson}, {Colafrancesco},
  {Cologna}, {Conrad}, {Couturier}, {Cui}, {Dalton}, {Daniel}, {Davids},
  {Degrange}, {Deil}, {deWilt}, {Dickinson}, {Djannati-Ata{\"\i}}, {Domainko},
  {Drury}, {Dubus}, {Dutson}, {Dyks}, {Dyrda}, {Edwards}, {Egberts}, {Eger},
  {Espigat}, {Farnier}, {Fegan}, {Feinstein}, {Fernandes}, {Fernandez},
  {Fiasson}, {Fontaine}, {F{\"o}rster}, {F{\"u}{\ss}ling}, {Gajdus}, {Gallant},
  {Garrigoux}, {Giavitto}, {Giebels}, {Glicenstein}, {Grondin},
  {Grudzi{\'n}ska}, {H{\"a}ffner}, {Hahn}, {Harris}, {Heinzelmann}, {Henri},
  {Hermann}, {Hervet}, {Hillert}, {Hinton}, {Hofmann}, {Hofverberg}, {Holler},
  {Horns}, {Jacholkowska}, {Jahn}, {Jamrozy}, {Janiak}, {Jankowsky}, {Jung},
  {Kastendieck}, {Katarzy{\'n}ski}, {Katz}, {Kaufmann}, {Kh{\'e}lifi},
  {Kieffer}, {Klepser}, {Klochkov}, {Klu{\'z}niak}, {Kneiske}, {Kolitzus},
  {Komin}, {Kosack}, {Krakau}, {Krayzel}, {Kr{\"u}ger}, {Laffon}, {Lamanna},
  {Lefaucheur}, {Lemi{\`e}re}, {Lemoine-Goumard}, {Lenain}, {Lennarz}, {Lohse},
  {Lopatin}, {Lu}, {Marandon}, {Marcowith}, {Marx}, {Maurin}, {Maxted},
  {Mayer}, {McComb}, {M{\'e}hault}, {Meintjes}, {Menzler}, {Meyer}, {Moderski},
  {Mohamed}, {Moulin}, {Murach}, {Naumann}, {de Naurois}, {Niemiec}, {Nolan},
  {Oakes}, {Ohm}, {de O{\~n}a Wilhelmi}, {Opitz}, {Ostrowski}, {Oya}, {Panter},
  {Parsons}, {Paz Arribas}, {Pekeur}, {Pelletier}, {Perez}, {Petrucci},
  {Peyaud}, {Pita}, {Poon}, {P{\"u}hlhofer}, {Punch}, {Quirrenbach}, {Raab},
  {Raue}, {Reimer}, {Reimer}, {Renaud}, {de los Reyes}, {Rieger}, {Rob},
  {Romoli}, {Rosier-Lees}, {Rowell}, {Rudak}, {Rulten}, {Sahakian}, {Sanchez},
  {Santangelo}, {Schlickeiser}, {Sch{\"u}ssler}, {Schulz}, {Schwanke},
  {Schwarzburg}, {Schwemmer}, {Sol}, {Spengler}, {Spies}, {Stawarz},
  {Steenkamp}, {Stegmann}, {Stinzing}, {Stycz}, {Sushch}, {Szostek},
  {Tavernet}, {Tavernier}, {Taylor}, {Terrier}, {Tluczykont}, {Trichard},
  {Valerius}, {van Eldik}, {van Soelen}, {Vasileiadis}, \&
  {Venter}}]{2015MNRAS.446.1163H}
{HESS Collaboration}, {Abramowski}, A., {Acero}, F., {et~al.} 2015, \mnras,
  446, 1163, \dodoi{10.1093/mnras/stu2148}

\bibitem[{{Hu} \& {Takata}(2021)}]{2021ApJ...922..260X}
{Hu}, X.~X., \& {Takata}, J. 2021, \apj, 922, 260,
  \dodoi{10.3847/1538-4357/ac273b}

\bibitem[{{Ignace} {et~al.}(2022){Ignace}, {Fullard}, {Shrestha}, {Naz{\'e}},
  {Gayley}, {Hoffman}, {Lomax}, \& {St-Louis}}]{2022ApJ...933....5I}
{Ignace}, R., {Fullard}, A., {Shrestha}, M., {et~al.} 2022, \apj, 933, 5,
  \dodoi{10.3847/1538-4357/ac6fce}

\bibitem[{{Jaron} {et~al.}(2024){Jaron}, {Kiehlmann}, \&
  {Readhead}}]{2024A&A...683A.228J}
{Jaron}, F., {Kiehlmann}, S., \& {Readhead}, A.~C.~S. 2024, \aap, 683, A228,
  \dodoi{10.1051/0004-6361/202347871}

\bibitem[{{Johnston} {et~al.}(1992){Johnston}, {Manchester}, {Lyne}, {Bailes},
  {Kaspi}, {Qiao}, \& {D'Amico}}]{1992ApJ...387L..37J}
{Johnston}, S., {Manchester}, R.~N., {Lyne}, A.~G., {et~al.} 1992, \apjl, 387,
  L37, \dodoi{10.1086/186300}

\bibitem[{{Jones} {et~al.}(2016){Jones}, {Halonen}, \&
  {Demers}}]{2016ASPC..506..165J}
{Jones}, C.~E., {Halonen}, R.~J., \& {Demers}, Z.~T. 2016, in Astronomical
  Society of the Pacific Conference Series, Vol. 506, Bright Emissaries: Be
  Stars as Messengers of Star-Disk Physics, ed. T.~A.~A. {Sigut} \& C.~E.
  {Jones}, 165

\bibitem[{{Kaaret} {et~al.}(2024){Kaaret}, {Roberts}, {Ehlert}, {Swartz},
  {Weisskopf}, {Liodakis}, {Saade}, {O'Dell}, \& {Chen}}]{2024ApJ...974L...1K}
{Kaaret}, P., {Roberts}, O.~J., {Ehlert}, S.~R., {et~al.} 2024, \apjl, 974, L1,
  \dodoi{10.3847/2041-8213/ad7ba6}

\bibitem[{{Kastner} \& {Mazzali}(1989)}]{1989A&A...210..295K}
{Kastner}, J.~H., \& {Mazzali}, P.~A. 1989, \aap, 210, 295

\bibitem[{{Kefala} \& {Bosch-Ramon}(2023)}]{2023A&A...669A..21K}
{Kefala}, E., \& {Bosch-Ramon}, V. 2023, \aap, 669, A21,
  \dodoi{10.1051/0004-6361/202244531}

\bibitem[{{Kirk} {et~al.}(1999){Kirk}, {Ball}, \&
  {Skj{\ae}raasen}}]{1999APh....10...31K}
{Kirk}, J.~G., {Ball}, L., \& {Skj{\ae}raasen}, O. 1999, Astroparticle Physics,
  10, 31, \dodoi{10.1016/S0927-6505(98)00041-3}

\bibitem[{{Kissmann} {et~al.}(2023){Kissmann}, {Huber}, \&
  {Gschwandtner}}]{2023A&A...677A...5K}
{Kissmann}, R., {Huber}, D., \& {Gschwandtner}, P. 2023, \aap, 677, A5,
  \dodoi{10.1051/0004-6361/202345934}

\bibitem[{{Kravtsov} {et~al.}(2020){Kravtsov}, {Berdyugin}, {Piirola},
  {Kosenkov}, {Tsygankov}, {Chernyakova}, {Malyshev}, {Sakanoi}, {Kagitani},
  {Berdyugina}, \& {Poutanen}}]{2020A&A...643A.170K}
{Kravtsov}, V., {Berdyugin}, A.~V., {Piirola}, V., {et~al.} 2020, \aap, 643,
  A170, \dodoi{10.1051/0004-6361/202038745}

\bibitem[{{Lamers} \& {Cassinelli}(1999)}]{1999isw..book.....L}
{Lamers}, H. J.~G.~L.~M., \& {Cassinelli}, J.~P. 1999, {Introduction to Stellar
  Winds}

\bibitem[{{LHAASO Collaboration}(2024)}]{2024arXiv241008988L}
{LHAASO Collaboration}. 2024, arXiv e-prints, arXiv:2410.08988,
  \dodoi{10.48550/arXiv.2410.08988}

\bibitem[{{Lyne} {et~al.}(2015){Lyne}, {Stappers}, {Keith}, {Ray}, {Kerr},
  {Camilo}, \& {Johnson}}]{2015MNRAS.451..581L}
{Lyne}, A.~G., {Stappers}, B.~W., {Keith}, M.~J., {et~al.} 2015, \mnras, 451,
  581, \dodoi{10.1093/mnras/stv236}

\bibitem[{{Manchester} {et~al.}(2005){Manchester}, {Hobbs}, {Teoh}, \&
  {Hobbs}}]{2005AJ....129.1993M}
{Manchester}, R.~N., {Hobbs}, G.~B., {Teoh}, A., \& {Hobbs}, M. 2005, \aj, 129,
  1993, \dodoi{10.1086/428488}

\bibitem[{{Maraschi} \& {Treves}(1981)}]{1981MNRAS.194P...1M}
{Maraschi}, L., \& {Treves}, A. 1981, \mnras, 194, 1P,
  \dodoi{10.1093/mnras/194.1.1P}

\bibitem[{{Massi} {et~al.}(2017){Massi}, {Migliari}, \&
  {Chernyakova}}]{2017MNRAS.468.3689M}
{Massi}, M., {Migliari}, S., \& {Chernyakova}, M. 2017, \mnras, 468, 3689,
  \dodoi{10.1093/mnras/stx778}

\bibitem[{{Massi} {et~al.}(2020){Massi}, {Chernyakova}, {Kraus}, {Malyshev},
  {Jaron}, {Kiehlmann}, {Dzib}, {Sharma}, {Migliari}, \&
  {Readhead}}]{2020MNRAS.498.3592M}
{Massi}, M., {Chernyakova}, M., {Kraus}, A., {et~al.} 2020, \mnras, 498, 3592,
  \dodoi{10.1093/mnras/staa2623}

\bibitem[{McSwain {et~al.}(2011)McSwain, Ray, Ransom, Roberts, Dougherty, \&
  Pooley}]{2011McSwainRadio}
McSwain, M.~V., Ray, P.~S., Ransom, S.~M., {et~al.} 2011, The Astrophysical
  Journal, 738, 105

\bibitem[{{Moritani} \& {Kawachi}(2021)}]{2021Univ....7..320M}
{Moritani}, Y., \& {Kawachi}, A. 2021, Universe, 7, 320,
  \dodoi{10.3390/universe7090320}

\bibitem[{{Moritani} {et~al.}(2015){Moritani}, {Okazaki}, {Carciofi}, {Imada},
  {Akitaya}, {Ebisuda}, {Itoh}, {Kawaguchi}, {Mori}, {Takaki}, {Ueno}, \&
  {Ui}}]{2015ApJ...804L..32M}
{Moritani}, Y., {Okazaki}, A.~T., {Carciofi}, A.~C., {et~al.} 2015, \apjl, 804,
  L32, \dodoi{10.1088/2041-8205/804/2/L32}

\bibitem[{{Nagae} {et~al.}(2006){Nagae}, {Kawabata}, {Fukazawa}, {Yamashita},
  {Ohsugi}, {Uemura}, {Chiyonobu}, {Isogai}, {Cho}, {Suzuki}, {Okazaki},
  {Okita}, \& {Yanagisawa}}]{2006PASJ...58.1015N}
{Nagae}, O., {Kawabata}, K.~S., {Fukazawa}, Y., {et~al.} 2006, \pasj, 58, 1015,
  \dodoi{10.1093/pasj/58.6.1015}

\bibitem[{{Okazaki} {et~al.}(2011){Okazaki}, {Nagataki}, {Naito}, {Kawachi},
  {Hayasaki}, {Owocki}, \& {Takata}}]{2011PASJ...63..893O}
{Okazaki}, A.~T., {Nagataki}, S., {Naito}, T., {et~al.} 2011, \pasj, 63, 893,
  \dodoi{10.1093/pasj/63.4.893}

\bibitem[{{Papitto} {et~al.}(2012){Papitto}, {Torres}, \&
  {Rea}}]{2012ApJ...756..188P}
{Papitto}, A., {Torres}, D.~F., \& {Rea}, N. 2012, \apj, 756, 188,
  \dodoi{10.1088/0004-637X/756/2/188}

\bibitem[{{Prinja}(1989)}]{1989Prinja}
{Prinja}, R.~K. 1989, \mnras, 241, 721, \dodoi{10.1093/mnras/241.4.721}

\bibitem[{{Prinja} {et~al.}(1990){Prinja}, {Barlow}, \&
  {Howarth}}]{1990ApJ...361..607P}
{Prinja}, R.~K., {Barlow}, M.~J., \& {Howarth}, I.~D. 1990, \apj, 361, 607,
  \dodoi{10.1086/169224}

\bibitem[{{Rivinius} {et~al.}(2013){Rivinius}, {Carciofi}, \&
  {Martayan}}]{2013A&ARv..21...69R}
{Rivinius}, T., {Carciofi}, A.~C., \& {Martayan}, C. 2013, \aapr, 21, 69,
  \dodoi{10.1007/s00159-013-0069-0}

\bibitem[{{R{\"u}bke} {et~al.}(2023){R{\"u}bke}, {Herrero}, \&
  {Puls}}]{2023A&A...679A..19R}
{R{\"u}bke}, K., {Herrero}, A., \& {Puls}, J. 2023, \aap, 679, A19,
  \dodoi{10.1051/0004-6361/202346487}

\bibitem[{{Runacres} \& {Owocki}(2002)}]{2002A&A...381.1015R}
{Runacres}, M.~C., \& {Owocki}, S.~P. 2002, \aap, 381, 1015,
  \dodoi{10.1051/0004-6361:20011526}

\bibitem[{{Saavedra} {et~al.}(2023){Saavedra}, {Romero}, {Bosch-Ramon}, \&
  {Kefala}}]{2023MNRAS.525.1848S}
{Saavedra}, E.~A., {Romero}, G.~E., {Bosch-Ramon}, V., \& {Kefala}, E. 2023,
  \mnras, 525, 1848, \dodoi{10.1093/mnras/stad2377}

\bibitem[{{Sathyaprakash} {et~al.}(2024){Sathyaprakash}, {Rea}, {Coti Zelati},
  {Borghese}, {Pilia}, {Trudu}, {Burgay}, {Turolla}, {Zane}, {Esposito},
  {Mereghetti}, {Campana}, {G{\"o}tz}, {Ibrahim}, {Israel}, {Possenti}, \&
  {Tiengo}}]{2024ApJ...976...56S}
{Sathyaprakash}, R., {Rea}, N., {Coti Zelati}, F., {et~al.} 2024, \apj, 976,
  56, \dodoi{10.3847/1538-4357/ad8226}

\bibitem[{{Snow}(1981)}]{1981Snow}
{Snow}, Jr., T.~P. 1981, \apj, 251, 139, \dodoi{10.1086/159448}

\bibitem[{{Suvorov} \& {Glampedakis}(2022)}]{2022ApJ...940..128S}
{Suvorov}, A.~G., \& {Glampedakis}, K. 2022, \apj, 940, 128,
  \dodoi{10.3847/1538-4357/ac9b48}

\bibitem[{{Takata} {et~al.}(2017){Takata}, {Tam}, {Ng}, {Li}, {Kong}, {Hui}, \&
  {Cheng}}]{2017ApJ...836..241T}
{Takata}, J., {Tam}, P.~H.~T., {Ng}, C.~W., {et~al.} 2017, \apj, 836, 241,
  \dodoi{10.3847/1538-4357/aa5c80}

\bibitem[{{Takata} {et~al.}(2012){Takata}, {Okazaki}, {Nagataki}, {Naito},
  {Kawachi}, {Lee}, {Mori}, {Hayasaki}, {Yamaguchi}, \&
  {Owocki}}]{2012ApJ...750...70T}
{Takata}, J., {Okazaki}, A.~T., {Nagataki}, S., {et~al.} 2012, \apj, 750, 70,
  \dodoi{10.1088/0004-637X/750/1/70}

\bibitem[{{Tavani} \& {Arons}(1997)}]{1997ApJ...477..439T}
{Tavani}, M., \& {Arons}, J. 1997, \apj, 477, 439, \dodoi{10.1086/303676}

\bibitem[{{Torres} {et~al.}(2012){Torres}, {Rea}, {Esposito}, {Li}, {Chen}, \&
  {Zhang}}]{2012ApJ...744..106T}
{Torres}, D.~F., {Rea}, N., {Esposito}, P., {et~al.} 2012, \apj, 744, 106,
  \dodoi{10.1088/0004-637X/744/2/106}

\bibitem[{{Waters} {et~al.}(1988){Waters}, {van den Heuvel}, {Taylor},
  {Habets}, \& {Persi}}]{1988Waters}
{Waters}, L.~B.~F.~M., {van den Heuvel}, E.~P.~J., {Taylor}, A.~R., {Habets},
  G.~M.~H.~J., \& {Persi}, P. 1988, \aap, 198, 200

\bibitem[{{Weisskopf} {et~al.}(2022){Weisskopf}, {Soffitta}, {Baldini},
  {Ramsey}, {O'Dell}, {Romani}, {Matt}, {Deininger}, {Baumgartner},
  {Bellazzini}, {Costa}, {Kolodziejczak}, {Latronico}, {Marshall}, {Muleri},
  {Bongiorno}, {Tennant}, {Bucciantini}, {Dovciak}, {Marin}, {Marscher},
  {Poutanen}, {Slane}, {Turolla}, {Kalinowski}, {Di Marco}, {Fabiani},
  {Minuti}, {La Monaca}, {Pinchera}, {Rankin}, {Sgro'}, {Trois}, {Xie},
  {Alexander}, {Allen}, {Amici}, {Andersen}, {Antonelli}, {Antoniak},
  {Attin{\`a}}, {Barbanera}, {Bachetti}, {Baggett}, {Bladt}, {Brez}, {Bonino},
  {Boree}, {Borotto}, {Breeding}, {Brienza}, {Bygott}, {Caporale}, {Cardelli},
  {Carpentiero}, {Castellano}, {Castronuovo}, {Cavalli}, {Cavazzuti},
  {Ceccanti}, {Centrone}, {Citraro}, {D'Amico}, {D'Alba}, {Di Gesu}, {Del
  Monte}, {Dietz}, {Di Lalla}, {Persio}, {Dolan}, {Donnarumma}, {Evangelista},
  {Ferrant}, {Ferrazzoli}, {Ferrie}, {Footdale}, {Forsyth}, {Foster},
  {Garelick}, {Gunji}, {Gurnee}, {Head}, {Hibbard}, {Johnson}, {Kelly},
  {Kilaru}, {Lefevre}, {Roy}, {Loffredo}, {Lorenzi}, {Lucchesi}, {Maddox},
  {Magazzu}, {Maldera}, {Manfreda}, {Mangraviti}, {Marengo}, {Marrocchesi},
  {Massaro}, {Mauger}, {McCracken}, {McEachen}, {Mize}, {Mereu}, {Mitchell},
  {Mitsuishi}, {Morbidini}, {Mosti}, {Nasimi}, {Negri}, {Negro}, {Nguyen},
  {Nitschke}, {Nuti}, {Onizuka}, {Oppedisano}, {Orsini}, {Osborne}, {Pacheco},
  {Paggi}, {Painter}, {Pavelitz}, {Pentz}, {Piazzolla}, {Perri},
  {Pesce-Rollins}, {Peterson}, {Pilia}, {Profeti}, {Puccetti}, {Ranganathan},
  {Ratheesh}, {Reedy}, {Root}, {Rubini}, {Ruswick}, {Sanchez}, {Sarra},
  {Santoli}, {Scalise}, {Sciortino}, {Schroeder}, {Seek}, {Sosdian}, {Spandre},
  {Speegle}, {Tamagawa}, {Tardiola}, {Tobia}, {Thomas}, {Valerie}, {Vimercati},
  {Walden}, {Weddendorf}, {Wedmore}, {Welch}, {Zanetti}, \&
  {Zanetti}}]{2022JATIS...8b6002W}
{Weisskopf}, M.~C., {Soffitta}, P., {Baldini}, L., {et~al.} 2022, Journal of
  Astronomical Telescopes, Instruments, and Systems, 8, 026002,
  \dodoi{10.1117/1.JATIS.8.2.026002}

\bibitem[{{Weng} {et~al.}(2022){Weng}, {Qian}, {Wang}, {Torres}, {Papitto},
  {Jiang}, {Xu}, {Li}, {Yan}, {Liu}, {Ge}, \& {Yuan}}]{2022NatAs...6..698W}
{Weng}, S.-S., {Qian}, L., {Wang}, B.-J., {et~al.} 2022, Nature Astronomy, 6,
  698, \dodoi{10.1038/s41550-022-01630-1}

\bibitem[{{Wood} {et~al.}(1996){Wood}, {Bjorkman}, {Whitney}, \&
  {Code}}]{1996ApJ...461..828W}
{Wood}, K., {Bjorkman}, J.~E., {Whitney}, B.~A., \& {Code}, A.~D. 1996, \apj,
  461, 828, \dodoi{10.1086/177105}

\bibitem[{{Yudin}(2014)}]{2014MNRAS.445.1761Y}
{Yudin}, R.~V. 2014, \mnras, 445, 1761, \dodoi{10.1093/mnras/stu1805}

\bibitem[{{Yudin} {et~al.}(2017){Yudin}, {Potter}, \&
  {Townsend}}]{2017MNRAS.464.4325Y}
{Yudin}, R.~V., {Potter}, S.~B., \& {Townsend}, L.~J. 2017, \mnras, 464, 4325,
  \dodoi{10.1093/mnras/stw2676}

\bibitem[{{Zabalza} {et~al.}(2013){Zabalza}, {Bosch-Ramon}, {Aharonian}, \&
  {Khangulyan}}]{2013A&A...551A..17Z}
{Zabalza}, V., {Bosch-Ramon}, V., {Aharonian}, F., \& {Khangulyan}, D. 2013,
  \aap, 551, A17, \dodoi{10.1051/0004-6361/201220589}

\bibitem[{{Zamanov} {et~al.}(2013){Zamanov}, {Stoyanov}, {Mart{\'\i}}, {Tomov},
  {Belcheva}, {Luque-Escamilla}, \& {Latev}}]{2013A&A...559A..87Z}
{Zamanov}, R., {Stoyanov}, K., {Mart{\'\i}}, J., {et~al.} 2013, \aap, 559, A87,
  \dodoi{10.1051/0004-6361/201321991}

\bibitem[{{Zhou} {et~al.}(2025){Zhou}, {Mao}, {Zhang}, {Patruno}, {Bozzo},
  {Xu}, {Santangelo}, {Zane}, {Zhang}, {Feng}, {Cavecchi}, {De Marco}, {Fan},
  {Hou}, {Jiang}, {Romano}, {Sala}, {Tao}, {Veledina}, {Vink}, {Wang}, {Wang},
  {Wang}, {Weng}, {Wu}, {Xie}, {Zhang}, {Zhang}, {Zhao}, {Zheng}, {Barua},
  {Chen}, {Chen}, {Chen}, {Chen}, {Chen}, {Cheng}, {Chi}, {Cui}, {de Martino},
  {Deng}, {Ducci}, {Farinelli}, {Feng}, {Ge}, {Gu}, {Guo}, {Han}, {Hu},
  {Huang}, {in't Zand}, {Ji}, {Kang}, {Kini}, {Li}, {Li}, {Liu}, {Liu}, {Liu},
  {Lyu}, {Marino}, {Markowitz}, {Mezcua}, {Middleton}, {Mou}, {Ng}, {Papitto},
  {Pei}, {Peng}, {Poutanen}, {Shui}, {Simone}, {Su}, {Tan}, {Wang}, {Wang},
  {Wang}, {Wang}, {Wang}, {Wang}, {Wang}, {Wu}, {Xiao}, {Xiong}, {Xu}, {Xue},
  {Yan}, {Yang}, {Yang}, {Yang}, {Ye}, {Yu}, {Yuan}, {Zhang}, {Zhang}, {Zhao},
  {Zhao}, {Zheng}, {Zheng}, \& {Zuo}}]{2025Zhou}
{Zhou}, P., {Mao}, J., {Zhang}, L., {et~al.} 2025, arXiv e-prints,
  arXiv:2506.08367, \dodoi{10.48550/arXiv.2506.08367}

\end{thebibliography}

\appendix
\section{Stokes parameters of the Thomson scattering}
\label{calculation}
\subsection{Basic equations}
In this section, we describe the method used to calculate the polarization resulting from Thomson scattering of the stellar wind, explicitly accounting for the finite size of the radiating star. We assume spherical symmetric structures for the stellar wind and star. For the basic process for deriving the Stokes parameters, we refer to \cite{1978A&A....68..415B}. We denote the unit vector of the coordinate "i" as $"\mathbf{e}_i"$, and introduce the Cartesian coordinate system ($x, y, z$), where the $x$ axis is the direction of the observer measured from the center of the star, and $z$ represents the orbital axis projected on the sky; please note that this coordinate system is different from the one illustrated in Figure~\ref{fig:psr}. We denote the spherical coordinate of a specific scattering point, $P$, as $(r,\theta,\phi)$. For the specific scattering point, $P$, we introduce another coordinate system, ($x', y', z'$), for which we take the direction of the $z'$ aligning with the radial direction to the scattering point from the center of the star, namely $\mathbf{e}_{z'}=\mathbf{e}_r$, $\mathbf{e}_{x'}=\mathbf{e}_\theta$, and $\mathbf{e}_{y'}=\mathbf{e}_\phi$. In this coordinate system, the emission point on the stellar surface, ($x'_e, y'_e, z'_e)$, is described by
\begin{equation}
x'_e=R_*\sin\theta'_e\cos\phi_e',~~y'_e=R_*\sin\theta'_e\sin\phi_e',~~z'_e=R_*\cos\theta'_e,
\end{equation}
where the polar coordinates $(\theta'_e, \phi'_e)$ represent the direction of the emission point measured from the direction of $z'$. This coordinate system is particularly useful for integrating the emissions over the stellar surface, since the integration rages are $0\le\theta'_e\le{\rm cos}^{-1}(R_*/r)$ and $0\le\phi'_e\le 2\pi$, respectively. Using the aforementioned two coordinate system, we will find the position vector of the scattering point measured from an emission point on the stellar surface as 
\begin{equation}
\mathbf{r}_{\rm ep}=x_{\rm ep}\mathbf{e}_x+y_{\rm ep}\mathbf{e}_y+z_{\rm ep}\mathbf{e}_z,~
\left\{\begin{array}{l}
x_{\rm ep}= (x-x'_e\cos\theta\cos\phi+y'_e\sin\phi-z'_e\sin\theta\cos\phi) \nonumber \\
y_{\rm ep}= (y-x'_e\cos\theta\sin\phi-y'_e\cos\phi-z'_e\sin\theta\sin\phi) \nonumber \\
z_{\rm ep}= (z+x'_e\sin\theta-z'_e\cos\theta).
\end{array}\right.
\label{conv}
\end{equation}.
We assume that the intensity from an emission point on the stellar surface received at the point $P$ is given by
\begin{equation}
    I(\theta_{\rm ep})=I_*\cos\theta_{\rm ep},
\end{equation}
where $\theta_{\rm ep}$ is the angle between the direction of the local surface normal and the direction to the point $P$ measured from the emission point. At the point $P$, the polarization caused by the Thomson scattering with the radiations integrated over the stellar surface may be described as~\citep{1978A&A....68..415B},
\begin{equation}
\left.  \begin{array}{c}
\triangle I(P)\\
\triangle Q(P) \\
\triangle U(P)
\end{array}\right\}
=\left[\frac{n(r)}{r^2}\triangle r\right]I_*\sigma_0\times\int\int\sin\theta_e'\cos\theta_{\rm ep}{\rm d}\theta_e'{\rm d}\phi_e' 
\left\{ \begin{array}{c}
1+\cos^2\chi_{\rm ep}\\
\sin^2\chi_{\rm ep}\cos2\phi_{\rm ep}\\
\sin^2\chi_{\rm ep}\sin2\phi_{\rm ep}
\end{array} \right.,
\end{equation}
where $n(r)$ is the electron number density, $I_*$ is the stellar intensity and $\sigma_0=3\sigma_T/16$ with $\sigma_T$ being the Thomson cross section. In addition, there is a relation that $\cos\chi_{\rm ep}=x_{ep}/r_{\rm ep}$, $\sin\chi_{\rm ep}\cos\phi_{\rm ep}=y_{\rm ep}/r_{\rm ep}$ and $\sin\chi_{\rm ep}\sin\phi_{\rm ep}=z_{\rm ep}$. Carrying out the integration over the stellar surface, we will obtain
\begin{equation}
  \left.  \begin{array}{c}
    \triangle I(P)\\
    \triangle Q(P) \\
    \triangle U(P)
  \end{array}\right\}
  =\pi\left[\frac{n(r)}{r^{2}}\triangle r\right]I_*\sigma_0\times
  \left\{ \begin{array}{l}
    D_0(r)+D_1(r)+D_2(r)\sin^2\theta\cos^2\phi\\
    D_2(r)(\sin^2\theta\sin^2\phi-\cos^2\theta)\\
    D_2(r)\sin2\theta\sin\phi
  \end{array} \right.,
  \label{eq:dstokes}
\end{equation}
where
\begin{eqnarray}
 D_0(r)&=&\frac{1}{\tilde{R}^2}\left[\tilde{t}^{1/2}\left(1-\tilde{R}^2-\frac{1}{3}\tilde{t}\right)\right]_0^{\theta'_{\rm e,max}} \nonumber \\
  D_{1}(r)&=&\frac{1}{8\tilde{R}^2}
  \left\{\tilde{t}^{-1/2}\left[\tilde{u}^3+(3+\tilde{R}^2)\tilde{u}\tilde{t}-\frac{1}{3}(3+\tilde{R}^2)\tilde{t}^2+\frac{1}{5}\tilde{t}^3\right]\right\}_0^{\theta'_{\rm e,max}}
  \nonumber \\
  D_{2}(r)&=&\frac{1}{8 \tilde{R}^2}\left\{\tilde{t}^{-1/2}\left[-3\tilde{u}^3-(1+3\tilde{R}^2)\tilde{u}\tilde{t}+\frac{1}{3}(1+3\tilde{R}^2)\tilde{t}^2-\frac{3}{5}\tilde{t}^{3}\right]\right\}_0^{\theta'_{\rm e,max}},
\end{eqnarray}
with $\theta'_{\rm e,max}={\rm cos}^{-1}(R_*/r)$, $\tilde{R}=R_*/r$, $\tilde{u}=1-\tilde{R}^2$, and $\tilde{t}(\theta'_e)=1+\tilde{R}^2-2\tilde{R}\cos\theta_e'$, which results in $\tilde{t}(0)=(1-\tilde{R})^2$ and $\tilde{t}(\theta'_{e,max})=\tilde{u}$. We integrate equation (\ref{eq:dstokes}) over the volume of the scattering region to obtain the Stokes parameters for comparisons with the observation data. In the limit of $\tilde{R}\rightarrow 0$ (point source limit), the coefficients approach to  $D_0\rightarrow 1$, $D_1\rightarrow 0$ and $D_2\rightarrow1$, recovering the equation~(3) of \cite{1978A&A....68..415B}.

As described in section~\ref{sec:assumption} and illustrated in Figure~\ref{fig:psr}, we assume that the geometry of the shock is axisymmetric structure about the line connecting between the pulsar and companion star. In such geometry of the scattering region, the Stokes parameters integrated over the scattering region can be rewritten as~\citep{1978A&A....68..415B},
\begin{equation}
\left.  \begin{array}{c}
\frac{I_i}{I_*}\\
\frac{Q_i}{I_*} \\
\frac{U_i}{I_*}
\end{array}\right\}
=\left\{ \begin{array}{l}
2\tau_{\rm c}+2(\tau_0 +\tau_0\gamma_0)+[\tau_0-3\tau_0\gamma_0 +\tau_0\gamma_3\cos(\nu-\nu_{\rm o})]\sin^2\theta_{\rm o}\\
(\tau_0-3\tau_0\gamma_0)\sin^2\theta_i-(1+\cos^2\theta_{\rm o})\tau_0\gamma_3\cos2(\nu-\nu_{\rm o})\\
-2\tau_0\gamma_3\cos\theta_o\sin2(\nu-\nu_{\rm o})
\end{array} \right.,
\label{eq:qui}
\end{equation}
where $\nu$ and $\nu_{\rm o}$ are the true anomaly of the pulsar and the observer, respectively (Figure~\ref{fig:orbital_cartoon}), and $\theta_{\rm o}$ is the direction of the observer measured from the orbital axis. In addition, we express that 
\begin{eqnarray}
 \tau_{\rm c}&=&\frac{\sigma_0}{2}\int\int\int \{n(D_0+D_1-D_2)\} \sin\Theta{\rm d}r{\rm d}\Theta
{\rm d}\Phi, \nonumber \\
\tau_0&=&\frac{\sigma_0}{2}\int\int \{nD_2\} \sin\Theta{\rm d}r{\rm d}\Theta{\rm d}\Phi, \nonumber \\
\tau_0\gamma_0&=&\frac{\sigma_0}{2}\int\int\int \{nD_2\sin^2\Theta\cos^2\Phi\} \sin\Theta{\rm d}r{\rm d}\Theta{\rm d}\Phi, \nonumber \\
\tau_0\gamma_3 &=& \frac{\sigma_0}{2}\int\int\int \{nD_2(\cos^2\Theta-\sin^2\Theta\sin^2\Phi)\}\sin\Theta{\rm d}r{\rm d}\Theta{\rm d}\Phi, 
\label{eq:taugamma}
\end{eqnarray}
where the polar coordinates $(\Theta,~\Phi)$ are measured about the line connecting the pulsar and the companion star, as illustrated in Figure~\ref{fig:psr}. We note that our definition of $(\Theta,~\Phi)$ are different from those used in equation~(7) of \cite{1978A&A....68..415B}.

\subsection{Thomson scattering from the shocked stellar wind}
In this study, we examine the structure of the shock explored by \cite{1996ApJ...469..729C}, which assume that the winds from two stars rapidly mix due to the radiative cooling. Using the conservations of mass and momentum, the column density ($\Sigma$) and the velocity ($v_{\rm s}$) of the shocked can be obtained. However, for the gamma-ray binary system, it is expected that radiation cooling is insufficient, and the contact discontinuity remains the scale of the binary system. Despite this, we adopt the shock model of \cite{1996ApJ...469..729C} as a simplified semi-analytical approach. A detailed treatment of the shock structure will require a numerical study~\citep[e.g.][]{2012ApJ...750...70T}

For the geometrically thin shock case, the integration of the radial direction in equation~(\ref{eq:taugamma}) may be replaced by 
\[
\int nD_2 dr\rightarrow \Sigma(r_{\rm s})D_2(r_{\rm s}), 
\]
where $r_{\rm s}(\Theta, \Phi)$ is the radial distance to a point of the shock surface from the center of the star, and $\Sigma(\Theta, \Phi)$ is the column density. The speed of the unshocked pulsar wind is $\sim c$, which is about two orders of magnitude faster than the unshocked stellar wind of $v_{\rm w}\sim 10^8~{\rm cm~s^{-1}}$. Moreover, the mass-loss rate of the stellar wind is much larger than that from the pulsar. Using these approximations, the velocity and the column mass density of the shocked stellar wind may be described by 
\begin{equation}
  \frac{v_{\rm s}(\Theta)}{v_{\rm w}}=\frac{\sqrt{[\eta(\Theta_1-\sin\Theta_1\cos\Theta_1)+(\Theta-\sin\Theta\cos\Theta)]^2+[\eta\sin^2\Theta_1-\sin^2\Theta]^2}}{2(1-\cos\Theta)}  
  \label{eq:vs}
\end{equation}
and
\begin{equation}
    \Sigma(\Theta)=\frac{\dot{M_1}}{2\pi v_{\rm w} r_{\rm s}(\Theta)}\times
    \frac{v_{\rm w}}{v_{\rm s}(\Theta)}\frac{\sin(\Theta+\Theta_1)(1-\cos\Theta)}{\sin\Theta\sin\Theta_1},
\end{equation}
where $\eta$ is the momentum ratio defined by equation~(\ref{eq:moment}) and $\Theta_1$, as illustrated in Figure~\ref{fig:psr}, is related to $\Theta$ as $\Theta{\rm cot}\Theta=1+\eta(\Theta_1{\rm cot}\Theta_1-1)$.

\section{inclination of the Be disk}
\label{sec:ibe}
This section we present how to determine the geometry of the Be disk from the results of the fitting for the polarization data. We express the direction of the Be disk axis as
\begin{equation}
n_{\rm disk}=\sin\theta_{\rm disk}\cos\nu_{\rm disk} e_{\rm x}+\sin\theta_{\rm disk}\sin\nu_{\rm disk} e_{\rm y}+\cos\theta_{\rm disk} e_{\rm z},
\end{equation}
where $\theta_{\rm disk}$ is measured from the orbit axis, and $\phi_{\rm disk}$ is measured from the direction of the periastron.  In this study, we assume that the direction of the linear polarization is aligned with the direction of the disk axis projected on the sky~\citep{2013Halonen}. Our study suggests that the contribution of the Be disk emission on P.D. is $\sim 1.7\%$ (Table~2). Such a high P.D. can be achieved if the Be disk is inclined from the observer with an angle of $\theta_{\rm o,d}\sim 70^{\circ}$~\citep{2013Halonen}, which is adopted in our study. Once $\theta_{\rm o,d}$ is given, we can deduce $\nu_{\rm disk}$ from the results of the fitting. Angle $\theta_{i,{\rm d}}$ can be expressed by 
\begin{equation}
    \cos\theta_{i,{\rm d}}=n_{\rm disk}\cdot n_{\rm o}= \sin\theta_{\rm o}\sin\theta_{\rm disk}\cos(\nu_{\rm disk}-\nu_{\rm o})+\cos\theta_{\rm o}\cos\theta_{\rm disk},
\label{eq:thi}
\end{equation}
where $\theta_{\rm o}$ is fixed to $\pi/3$~rad in this study and $\nu_{\rm o}$ is determined by the fitting process. From the definition of $\chi_0$ and $\chi_{\rm disk}$ of the P.A., we can obtain 
\begin{eqnarray}
  \cos(\chi_0-\chi_{\rm disk})&=&\frac{\sin\theta_{\rm disk}\sin(\nu_{\rm disk}-\nu_{\rm o})}{\sin\theta_{\rm o,d}} \nonumber \\
  \sin(\chi_0-\chi_{\rm disk})&=&\frac{\cos\theta_{\rm o}\sin\theta_{\rm disk}\cos(\nu_{\rm disk}-\nu_{\rm o})-\sin\theta_{\rm o}\cos\theta_{\rm disk}}{\sin\theta_{\rm o,d}}.
  \label{eq:cdis}
\end{eqnarray}
Using equations~(\ref{eq:thi}) and~(\ref{eq:cdis}) with the fitting results, we calculate $\theta_{\rm disk}$ and $\nu_{\rm disk}$. The node between the disk plane and the orbital plane is 
\begin{eqnarray}
    \nu_{\rm node}=\nu_{\rm disk}+\pi/2.
\end{eqnarray}

\end{document}